\documentclass[aps,prd,showpacs]{revtex4}
\usepackage[usenames]{color}
\usepackage[dvips]{epsfig}
\usepackage{citesort}
\usepackage{mathrsfs}
\usepackage{amsmath,amssymb}
\usepackage{dcolumn}% Align table columns on decimal point
\usepackage{bm}% bold math
\usepackage{tabularx}
\usepackage{feynmp}

\newcommand{\be}{\begin{equation}}
\newcommand{\ee}{\end{equation}}
\newcommand{\bea}{\begin{eqnarray}}
\newcommand{\eea}{\end{eqnarray}}
\newcommand{\bean}{\begin{eqnarray*}}
\newcommand{\eean}{\end{eqnarray*}}
\newcommand{\ba}{\begin{array}}
\newcommand{\ea}{\end{array}}
\newcommand{\non}{\nonumber}
\newcommand{\bc}{\begin{center}}
\newcommand{\ec}{\end{center}}
\newcommand{\bi}{\begin{itemize}}
\newcommand{\ei}{\end{itemize}}

\newcommand{\Slash}[1]{{\ooalign{\hfil/\hfil\crcr$#1$}}}

\newcommand{\bef}{\begin{figure}[htb]\centering}
\newcommand{\eef}{\end{figure}}
\newif\ifContLineOne
\newif\ifContLineTwo
\newif\ifContLineThree
\def\conC#1{\vbox{\ialign{##\crcr
  \ifContLineThree\hrulefill\else\vphantom{\hrulefill}\fi\crcr
  \noalign{\kern3.2pt\nointerlineskip}
  \ifContLineTwo\hrulefill\else\vphantom{\hrulefill}\fi\crcr
  \noalign{\kern3.2pt\nointerlineskip}
  \ifContLineOne\hrulefill\else\vphantom{\hrulefill}\fi\crcr
  \noalign{\nointerlineskip}
  $\hfil\textstyle{\vbox to 14pt{}#1}\hfil$\crcr}}}
\def\DrawLeg#1#2{
  \kern-.2pt              % back up half width of leg
  \dimen2 =#1             % =height of whatever is underneath leg
  \advance\dimen2 by 2pt  % 2pt space below bottom of leg
  \dimen3 = 10.6pt        % base value of height of top of leg
  \dimen4 =3.6pt          % add this much time 1 2 or 3 to base value
  \advance\dimen3 by -\dimen2 
  \multiply\dimen4 by #2
  \advance\dimen3 by \dimen4
  \raise\dimen2 \hbox{\vrule height\dimen3 width .4pt} % draw it
  \kern-.2pt}             % and back up half width of line
\def\begC#1#2{\setbox0 =\hbox{$\textstyle{#2}$}
  \dimen0=.5\wd0 \dimen1=\ht0
  \conC{\hskip\dimen0}
  \count255=#1
  \ifnum\count255 =1 \ContLineOnetrue\else
  \ifnum\count255 =2 \ContLineTwotrue\else
  \ifnum\count255 =3 \ContLineThreetrue\fi\fi\fi
  \DrawLeg{\dimen1}{\count255}
  \conC{\hskip\dimen0}
  \kern-\dimen0\kern-\dimen0 \box0}
\def\endC#1#2{\setbox0 =\hbox{$\textstyle{#2}$}
  \dimen0=.5\wd0 \dimen1=\ht0
  \conC{\hskip\dimen0}
  \count255=#1
  \ifnum\count255 =1 \ContLineOnefalse\else
  \ifnum\count255 =2 \ContLineTwofalse\else
  \ifnum\count255 =3 \ContLineThreefalse\fi\fi\fi
  \DrawLeg{\dimen1}{\count255}
  \conC{\hskip\dimen0}
  \kern-\dimen0\kern-\dimen0 \box0}
\begin{document}
\preprint{}
%% Title + Authors 
\title{Evolution equation for the $B$-meson distribution amplitude\\ 
in the heavy-quark effective theory in coordinate space}   

\author{Hiroyuki Kawamura$^{1}$ and Kazuhiro Tanaka$^2$}
\affiliation{${}^1$Department of Mathematical Sciences, University of Liverpool,
           Liverpool, L69 3BX, United Kingdom\\
${}^2$Department of Physics, Juntendo University, Inba-gun, Chiba
270-1695, Japan}
\date{\today}

%% Abstract        
\begin{abstract} 
The $B$-meson distribution amplitude (DA) is
defined as the matrix element 
of a quark-antiquark bilocal light-cone operator in the heavy-quark effective theory,
corresponding to a long-distance component
in the factorization formula for exclusive $B$-meson decays.
The evolution equation for the $B$-meson DA
is governed by the cusp anomalous dimension as well as 
the Dokshitzer-Gribov-Lipatov-Altarelli-Parisi-type anomalous dimension,
and these anomalous dimensions 
give the ``quasilocal'' kernel in the coordinate-space representation.
We show that this evolution equation can be solved analytically in the coordinate-space,
accomplishing the relevant Sudakov resummation at the next-to-leading logarithmic accuracy.
The quasilocal nature leads to a quite simple form of our solution
which determines the $B$-meson DA 
with a quark-antiquark light-cone separation $t$ 
in terms of the DA at a lower renormalization scale $\mu$ 
with smaller interquark separations $zt$ ($z\leq 1$).
This formula allows us to present 
rigorous calculation of the $B$-meson DA
at the factorization scale $\sim \sqrt{m_b \Lambda_{\rm QCD}}$ for $t$ less than $\sim 1$~GeV$^{-1}$,
using the recently obtained operator product 
expansion of the DA as the input at $\mu \sim 1$~GeV.
We also derive the master formula,
which reexpresses the integrals of the DA
at $\mu\sim\sqrt{m_b \Lambda_{\rm QCD}}$ for the factorization formula 
by the compact integrals of the DA at $\mu \sim 1$~GeV.
\end{abstract}
\pacs{12.38.Cy, 12.39.Hg, 12.39.St, 13.25.Hw}
%\keywords{}
\maketitle
%
%\newpage       
%\pagestyle{plain}
%\setcounter{page}{1}
%\setcounter{equation}{0}
%\renewcommand{\theequation}{\arabic{equation}}
%
%\vskip 1cm
%----------------------------------------------------------------------------------
\section{Introduction}
\label{sec1}
The $B$-meson light-cone distribution amplitude (LCDA) 
is one of the important
ingredients of the QCD factorization formula for exclusive 
$B$ decays~\cite{Beneke:2000ry,Bauer:2001cu,Li:2003yj}
and  
%the $B$-meson LCDA 
has recently attracted much attention due to its 
%importance 
central role for the analysis of the experimental data, e.g., hadronic and radiative
$B$-decay data~\cite{Antonelli:2009ws}.
The $B$-meson LCDA 
appears in the 
factorization formula for 
hard spectator interaction amplitudes, where a large momentum is 
transferred to the spectator light-quark via gluon 
exchange~\cite{Beneke:2000ry,Bauer:2001cu,Li:2003yj,Antonelli:2009ws,Korchemsky:1999qb,BBNL,BBNL2,Beneke:2000wa,Beneke:2005vv,Bell08},
and represents
the nonperturbative matrix element
that describes the leading amplitude to have 
the valence quark and antiquark 
with a light-like separation 
inside the $B$ meson~\cite{Szczepaniak:1990dt}.
Grozin and Neubert~\cite{Grozin:1997pq} studied
constraints on the $B$-meson LCDA from the equations of motion, heavy-quark symmetry and the 
renormalization group (RG), and they gave
the first quantitative estimate of the LCDA
using QCD sum rules with the leading perturbative and nonperturbative effects
taken into account.
The light-cone QCD sum rules for the $B$-decay form factors 
were also used~\cite{Ball:2003fq,Khodjamirian:2005ea} to estimate 
the first inverse moment~\cite{foot1}
%\footnote{
%%Here and below, 
%Unless otherwise indicated,
%the ``moment'' implies
%the one with respect to the momentum variable.}
%calculated 
%defined in the momentum representation of the LCDA.
of the LCDA, 
which participates in the corresponding factorization formula.
The Grozin-Neubert's QCD sum rule calculation
was extended by Braun, Ivanov and Korchemsky~\cite{Braun:2003wx} 
including the perturbative and nonperturbative corrections,
and the importance of the NLO perturbative corrections was emphasized.
Indeed, the true non-analytic behavior of the $B$-meson LCDA 
associated with the ``cusp singularities''~\cite{Korchemsky:1987wg}
% at short distance
is only revealed at this level including the radiative corrections~\cite{Lange:2003ff},
and it is this behavior 
%at short-distance 
that renders the (non-negative) moments of the LCDA divergent, 
even after renormalization~\cite{Grozin:1997pq} (see \cite{Ball:2008fw}
for a similar behavior in three-quark LCDAs for the $\Lambda_b$ baryon).
Introducing the regularization for the moments 
%of the LCDA 
with an additional momentum cutoff,
Lee and Neubert~\cite{Lee:2005gza} 
% $\Lambda_{UV}$,
evaluated 
the first two moments for a large value of the cutoff
in terms of
%using
the operator product expansion (OPE)
with the NLO perturbative corrections, as well as 
the power corrections which are
generated by
the local operators of dimension 4,
and they used the results as constrains on model building of the $B$-meson LCDA.

Another 
%remarkable 
feature peculiar to the $B$-meson LCDA
is that it involves a complicated mixture
of the multiparticle Fock states of higher-twist nature 
through nonperturbative quark-gluon interactions,
%with the nonperturbative gluons,
as demonstrated using the equations of motion and heavy-quark symmetry~\cite{Grozin:1997pq,KKQT,GW}.
%(see also \cite{GW}). 
A first systematic treatment of the mixing of the multiparticle states, 
disentangling the singularities
from the radiative corrections,
has recently been accomplished by the present authors,
and the $B$-meson LCDA is obtained 
in a form of the OPE as the short-distance expansion 
for the quark-antiquark light-cone separation,
with the subleading and subsubleading
%nonleading
%leading and next-to-leading 
power corrections, generated by
the local operators of dimension $d = 4$ and 5, 
respectively, 
and the NLO corrections
for the corresponding Wilson coefficients~\cite{Kawamura:2008vq}.
% in the $\overline{\rm MS}$ scheme, 
This OPE enables us to evaluate the $B$-meson LCDA 
%express 
%$\tilde{\phi}_+(t,\mu)$ 
for interquark distances $t$ with $t \lesssim 1/\mu$,
%less than $\sim 1/\mu$ with 
where $\mu$ is the renormalization scale of the LCDA, 
%$\sim 1$~GeV$^{-1}$, 
in a rigorous way
in terms of three nonperturbative parameters in the heavy-quark effective theory (HQET),
one of which is the usual mass difference between the $B$-meson and $b$-quark,
%HQET parameter, 
$\bar{\Lambda}=m_B -m_b$,
%the mass difference between the $B$-meson and $b$-quark
associated with matrix elements of
% as the matrix element
%of the 
dimension-4 operators,
and the other two are the novel HQET parameters~\cite{Grozin:1997pq,KKQT,GW} 
associated with matrix elements of the quark-antiquark-gluon three-body operators of dimension 5.
Note that the range of $t$ where the OPE is directly applicable 
becomes wider for  
the smaller value of the scale $\mu$, as $t \lesssim 1/\mu$:
choosing 
$\mu=1$~GeV, corresponding to typical hadronic scale, 
the model-independent result 
for interquark distances $t\lesssim 1$~GeV$^{-1}$ has been obtained from the OPE 
%(\ref{OPE}) 
and this result
has also been used to constrain the 
behavior of the LCDA 
at large distances $t \gtrsim 1$~GeV$^{-1}$~\cite{Kawamura:2008vq}.
Those results of the $B$-meson LCDA at $\mu=1$~GeV
have to be evolved to the factorization
scale 
%$\mu_F$ 
of order $\mu_{\rm hc} \sim \sqrt{m_b\Lambda_{\rm QCD}}$
that corresponds to the characteristic 
``hard-collinear'' scale
for hard spectator scattering 
in exclusive $B$ decays~\cite{Beneke:2000ry,Bauer:2001cu,Li:2003yj},
when we substitute the LCDA
into the relevant factorization formula.

For this purpose, in principle, we can utilize the analytic solution 
for the evolution equation of the $B$-meson LCDA
obtained in \cite{Lange:2003ff,Lee:2005gza}.
However, the corresponding solution is directly applicable
when the LCDA is given in the momentum representation, which we find 
inconvenient in our case: 
the Fourier transformation of the above OPE-based results 
to the momentum space
mixes up the model-independent behavior for $t\lesssim 1$~GeV$^{-1}$ 
with the behavior for $t \gtrsim 1$~GeV$^{-1}$ which relies on a certain model 
for the large $t$ behavior.
On the other hand, it has been noted that the relevant evolution kernel embodies the particularly simple
geometrical structure in the coordinate-space representation~\cite{Braun:2003wx}.
These facts motivate us to treat the evolution of the $B$-meson LCDA
in an unconventional way, working in the coordinate-space representation.
% in the present paper.
We are able to find the analytic solution for the corresponding evolution
equation, and demonstrate that the solution
determines
%the evolution of 
the $B$-meson LCDA
in terms of the LCDA at a lower scale $\mu$ 
with
smaller 
interquark 
separations and thus 
preserves the boundary at $t\sim 1$~GeV$^{-1}$ between
the model-independent and -dependent behaviors of our LCDA,
% of our prediction for $t\lesssim 1$~GeV$^{-1}$ 
even after evolving from 
%the hadronic scale 
$\mu=1$~GeV to 
%the hard-collinear scale 
$\mu_{\rm hc}$.
We emphasize that such simple RG structure of the $B$-meson LCDA 
can be manifested only in the coordinate space.
Furthermore, as we shall demonstrate, it is this simple structure that enables us to 
derive the master formula, by which
the relevant integrals of the LCDA
at the scale $\mu_{\rm hc}$, arising in 
the factorization formula for the exclusive $B$-meson decays,
can be reexpressed 
%in a model-independent way
by the compact integrals of the LCDA at the scale $\mu= 1$~GeV.
Therefore, we believe that the coordinate-space approach for 
the RG evolution of the $B$-meson LCDA deserves detailed discussions
in the present paper.
We also show that our solution can be organized so as to 
include the Sudakov resummation to the next-to-leading logarithmic (NLL) accuracy,
taking into account the effects of the anomalous dimension at the two-loop level which is associated 
with the cusp singularity.
We present the first rigorous result of
%first 
the $B$-meson LCDA
at the relevant factorization 
scale $\mu_{\rm hc}$ for $t\lesssim 1$~GeV$^{-1}$.
Combining with the results for the long-distance behavior,
% for $t\lesssim 1$~GeV$^{-1}$with 
we discuss an estimate for the inverse moments of the LCDA 
%the integral relevant to exclusive $B$ decays, $\lambda_B$,
at $\mu=\mu_{\rm hc}$. 

The paper is organized as follows.
Sec.~\ref{sec2} is mainly introductory; 
we give 
%the HQET light-cone bilocal operator 
the operator definition
of the $B$-meson LCDA, explain
the result 
for its renormalization 
%at one-loop level 
in the coordinate space,
and derive the corresponding RG evolution equation.
We demonstrate in Sec.~\ref{sec3} that, as the solution of this equation,
we can obtain the new coordinate-space representation 
for the evolution of the $B$-meson LCDA, which manifests the simple
RG structure, and also organize the result so as to include the Sudakov resummation at the NLL-level.
In Sec.~\ref{sec4}, we use
our coordinate-space representation of the evolution 
%allows us 
to derive a compact and closed formula for
the inverse moments of the LCDA 
%at the scale $\mu \sim \mu_{\rm hc}$
in terms of the certain integrals of the LCDA at a lower scale $\mu$.
Application of our results to calculate the evolution of the OPE-based form of the $B$-meson LCDA is presented in Sec.~\ref{sec5}, 
and we discuss an estimate for the inverse moments of the LCDA.
Sec.~\ref{sec6} is reserved for conclusions.

\section{Definition and renormalization in the coordinate space}
\label{sec2}

The leading quark-antiquark component of the $B$-meson LCDA is defined 
as the vacuum-to-meson matrix element in the HQET~\cite{Grozin:1997pq}:
\begin{equation}
\tilde{\phi}_+(t, \mu)
=\frac{1}{iF(\mu)}
\langle 0|
\bar{q}(tn)
[tn,0]
\Slash{n}
\gamma_5h_v(0) 
|\bar{B}(v)\rangle 
=\int d\omega e^{-i\omega t}
\phi_+(\omega, \mu)\ ,
\label{eq1}
\end{equation}
where $\bar{q}(tn)$ is the light-antiquark field, $h_v(0)$ is the effective 
heavy-quark field,
and these fields form a gauge-invariant bilocal operator linked by a light-like Wilson line,
\begin{equation}
[tn,0]={\cal P}\exp \left[ ig\int_0^td\lambda\ n\cdot A(\lambda n)\right]\ , 
\end{equation}
with $n_\mu$ as the light-like vector, $n^2 =0$ and $n\cdot v=1$,
and $v^\mu$ denoting the 4-velocity of the $B$ meson.  
The bilocal operator is renormalized at the scale $\mu$ and,
here and below, $\mu$ refers to the $\overline{\rm MS}$ renormalization scale.
In the definition~(\ref{eq1}), 
\begin{equation}
F(\mu)=-i\langle 0|\bar{q}\Slash{n}\gamma_5h_v |\bar{B}(v)\rangle
\label{fmu}
\end{equation}
denotes the $B$-meson decay constant in the HQET~\cite{Neubert:1994mb}
and $\phi_+(\omega, \mu)$ in the RHS is the LCDA in the momentum representation where $\omega v^+$ 
denotes the light-cone ``$+$''-component of the momentum carried by the light antiquark. 

The renormalization of the bilocal operator of (\ref{eq1})
was studied in \cite{Grozin:1997pq,Lange:2003ff},
calculating the UV divergence 
in the one-loop diagrams of Fig.~1 in the momentum space 
(see also \cite{Grozin:2005iz,DescotesGenon:2009hk}).
The calculation of those diagrams has also been carried out 
in the coordinate space~\cite{Braun:2003wx,Kawamura:2008vq}, and the result
yields the renormalization of the bilocal operator 
$\Theta(t)\equiv \bar{q}(tn)[tn,0]\Slash{n}\gamma_5h_v(0)$
in the coordinate-space representation as
(unless otherwise indicated, $\alpha_s \equiv \alpha_s (\mu)$)
\bea
\Theta^{\rm bare}(t)
&& \!\!\!\!\!\!
=\Theta^{\rm ren}(t, \mu)
+\frac{\alpha_sC_F}{2\pi}
%\int_0^1d\xi
\left\{\left(-\frac{1}{2\varepsilon^2}-\frac{L}{\varepsilon}
+\frac{1}{4\varepsilon} 
\right) 
\Theta^{\rm ren}(t, \mu)
\right.
\nonumber\\
&& \left.
+\frac{1}{\varepsilon} 
\int_0^1dz
\left(\frac{z}{1-z}\right)_+ \Theta^{\rm ren}(z t, \mu)
%+\frac{1}{4\varepsilon} \del(1-u)
\right\}\ ,
%{\cal O}_+(u t)  \ ,
\label{ren1}
\eea
connecting the bare and renormalized operators by the
``$z$-dependent'' renormalization constant
in $D=4-2\varepsilon$ dimensions and Feynman gauge,
where $C_F = (N_c^2-1)/(2N_c)$, and
\be
L= \ln\left[i(t-i0) \mu e^{\gamma_E}\right]\ ,
\label{L}
\ee
with 
%the ${\overline{\rm MS}}$ scale $\mu$ and 
the Euler constant $\gamma_E$ and
%Note that 
the ``$-i0$'' prescription coming from the position of the pole in 
the relevant propagators in the coordinate space. The plus-distribution 
%$[u/(1-u)]_+$ 
is defined, as usual, as
%for the interval from $0$ and $1$. 
\be
\int_0^1 dz \left(\frac{z}{1-z}\right)_+ f(z)
\equiv \int_0^1 dz \frac{z \left[f(z)-f(1)\right]}{1-z}\ ,
\label{plusd}
\ee
for a smooth test function $f(z)$.
In the one-loop contributions in (\ref{ren1}), the first two terms,
the double-pole term and the single-pole term involving $L$,
%The logarithmic term on the RHS 
manifest the cusp singularity~\cite{Korchemsky:1987wg} in Fig.~1~(a),
%in 
i.e., the singularity in 
the radiative correction around the cusp (at the origin) in the Wilson line,
\be
[tn,0] [0,-\infty v ]\ ,
\label{wl}
\ee
%between the two Wilson lines 
which is contained in (\ref{eq1}),
%, $[tn,0] [0,-\infty v ]$
using
the relation $h_v(0)= [0,-\infty v ] h_v(-\infty v)$.
The last term in (\ref{ren1}) comes from Fig.~1~(b),
accompanying the plus-distribution 
characteristic of the loop integral 
associated with the massless degrees of freedom only, 
while 
the remaining one-loop term
comes from the contribution of the renormalization constants 
of the two quark fields, $\bar{q}$ and $h_v$.
We note that
%Finally, the ladder-type diagram in 
Fig.~1~(c) is UV-finite in the Feynman gauge~\cite{foot2}
%\footnote{Figure~1~(a) is IR-finite,
%while Figs.~1~(b),~(c) produce the IR poles in the Feynman gauge. See \cite{Kawamura:2008vq}
%for the details.} 
and does not contribute to (\ref{ren1}).
%%%%%%%%%%%%%%%%%%%%%%%%%%%%%%%%%%%%%%%%%%%%%%%%%%%%%%%%%%%%%%%%%%%%%%%%%%%%%%%
\begin{figure}
\bc
\includegraphics[height=3cm,clip]{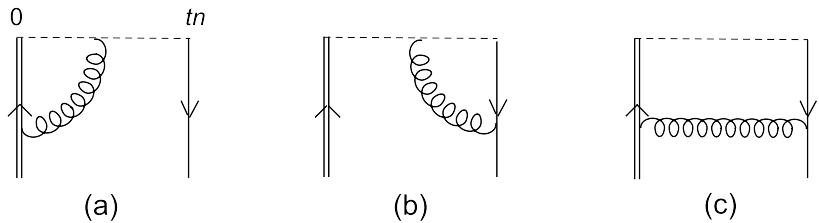}~~~~
\ec
\caption{
The Feynman diagrams relevant for the one-loop renormalization 
of the nonlocal light-cone operator 
in (\ref{eq1}). 
The dashed line represents the Wilson line in between the quark fields,
and the double line represents the effective heavy-quark field.}
\label{fig:1}
\end{figure}
%%%%%%%%%%%%%%%%%%%%%%%%%%%%%%%%%%%%%%%%%%%%%%%%%%%%%%%%%%%%%%%%%%%%%%%%%%%%%%%

The RG invariance to the one-loop accuracy based on (\ref{ren1})
implies that the $B$-meson LCDA (\ref{eq1})
obeys the evolution equation in the coordinate space,
\bea
\mu\frac{d}{d\mu}\tilde{\phi}_+(t,\mu)=
%\biggl\{-\biggl[
-\left[\Gamma_{\rm cusp}(\alpha_s)L + \gamma_F(\alpha_s)\right]
\tilde{\phi}_+(t,\mu)
%\biggr]\delta(1-u)
+
\int_0^1dz
K(z, \alpha_s) \tilde{\phi}_+(z t,\mu)\ ,
\label{evol}
\eea
with the one-loop RG functions 
\bea
\Gamma_{\rm cusp}(\alpha_s)&& \!\!\!\!\!\! =
\Gamma_{\rm cusp}^{(1)} \frac{\alpha_s}{4\pi}\ , \;\;\;\;\;\;\;\;\;\;
\Gamma_{\rm cusp}^{(1)} =4C_F\ , 
\label{rgf1}\\
\gamma_F(\alpha_s) && \!\!\!\!\!\! = \gamma^{(1)}_F \frac{\alpha_s}{4\pi}\ , 
\;\;\;\;\;\;\;\;\;\;
\gamma^{(1)}_F =2C_F\ , 
\label{rgf12}
\eea
and 
\be
K(z,\alpha_s)
= K^{(1)}(z)\frac{\alpha_s}{4\pi}\ ,
\;\;\;\;\;\;\;\;\;\;
K^{(1)}(z)=4C_F\left(\frac{z}{1-z}\right)_+\ .
\label{rgf2}
\ee
Here, $\Gamma_{\rm cusp}(\alpha_s)$ corresponds to the anomalous dimension
of the Wilson line with a cusp, (\ref{wl}),
and coincides with  
the LO
%order $\alpha_s$ 
term of 
the universal cusp anomalous dimension 
of Wilson loops with light-like segments~\cite{Korchemsky:1987wg}; 
we obtain the finite result (\ref{rgf1}),
because the contribution from the double-pole term of (\ref{ren1})
is canceled by the contribution generated by taking the derivative of $L$ in the next term
with respect to $\mu$.
$\gamma_F(\alpha_s)$ of (\ref{rgf12}) 
represents the anomalous dimension from the above-mentioned contribution of the
renormalization constants
of the
two quark fields, combined with the so-called hybrid 
anomalous dimension of heavy-light currents in the HQET~\cite{Neubert:1994mb},
which governs the scale dependence of
the decay constant (\ref{fmu}) as,
at one-loop accuracy,
%$-3C_F \alpha_s /(4\pi)$ ~\cite{Neubert:1994mb}.
\be
\mu\frac{d}{d\mu}F(\mu)= 3C_F \frac{\alpha_s}{4\pi}F(\mu)\ .
\label{evol:F}
\ee 
$K(z,\alpha_s)$ of (\ref{rgf2}) comes from the last term of (\ref{ren1})
and represents the $z$-dependent anomalous dimension associated with
the massless degrees of freedom only.
We note
the remarkable property in (\ref{evol}) that the evolution kernel in the RHS, composed of
(\ref{rgf1})-(\ref{rgf2}), is quasilocal,
such that
the evolution mixes the LCDA with itself and with the LCDA associated with smaller light-cone
separation $z t$ ($z <1$).
This is due to the similar structure 
appearing in the renormalization 
in the coordinate space, (\ref{ren1}), 
and reflects~\cite{Braun:2003wx} 
the fact
that the cusp renormalization induced by Fig.~1~(a) is multiplicative 
in the coordinate space~\cite{Korchemsky:1987wg} 
while Fig.~1~(b) gives the contribution identical to the similar correction
to the light-quark-antiquark bilocal operators, which 
embodies simple geometrical structure in the coordinate space 
so as to mix operators associated with smaller ``size''~\cite{Balitsky:1987bk}.

It is straightforward to perform the Fourier transformation of (\ref{evol}) 
to the momentum space
and derive the 
evolution equation for $\phi_+(\omega, \mu)$ of (\ref{eq1}),
using
\bea
\lefteqn{\frac{1}{2\pi}\int_{-\infty}^{\infty}dt e^{i\omega t}L\tilde{\phi}_+(t,\mu)}
\nonumber\\
&&=\frac{i}{2\pi}\int_{0}^\infty d\omega' \left(
\frac{1}{\omega'-\omega-i0}\ln\frac{\omega'-\omega -i0}{\mu}
-\frac{1}{\omega'-\omega+i0}\ln\frac{\omega'-\omega +i0}{\mu}
\right) \phi_+(\omega', \mu)
\nonumber\\
&&
=- \phi_+(\omega, \mu)\ln\frac{\omega}{\mu}- \int_0^\infty d\omega' 
\frac{\theta(\omega-\omega')}{\omega-\omega'}
\left[\phi_+(\omega', \mu)-\phi_+(\omega, \mu)\right]\ ,
\label{ft1}
\eea
and
\bea
\lefteqn{
\frac{1}{2\pi}\int_{-\infty}^{\infty}dt e^{i\omega t}
\int_0^1 dz \left(\frac{z}{1-z}\right)_+ \tilde{\phi}_+(z t,\mu)}
\nonumber\\
&&
=\phi_+(\omega, \mu) + \int_0^\infty d\omega' 
\frac{\omega}{\omega'}\frac{\theta(\omega'-\omega)}{\omega'-\omega}
\left[\phi_+(\omega', \mu)-\phi_+(\omega, \mu)\right]\ ,
\label{ft2}
\eea
and, indeed, the result 
reproduces the evolution equation obtained through the renormalization
of the bilocal operator of (\ref{eq1})
in the momentum space~\cite{Lange:2003ff,Grozin:2005iz,DescotesGenon:2009hk}.
We note that the momentum representation of the kernel in (\ref{ft2}) 
coincides with (a part of) 
the Brodsky-Lepage kernel
for the pion LCDA~\cite{Lepage:1979zb}
and physically represents 
%has a physical interpretation as 
a Dokshitzer-Gribov-Lipatov-Altarelli-Parisi (DGLAP) splitting function
that vanishes for $\omega/\omega' >1$.
On the other hand, 
%(\ref{ft1}) shows that 
a local contribution in (\ref{evol}), associated with the cusp anomalous 
dimension $\Gamma_{\rm cusp}(\alpha_s)$,
yields the new evolution kernel for $\omega/\omega' \ge 1$ 
%in the momemtum space, as shown 
in the RHS of (\ref{ft1}),
so that the evolution in the momentum space mixes the LCDA $\phi_{+}(\omega, \mu)$
%with itself and 
with $\phi_{+}(\omega', \mu)$ over
the entire region, 
$0< \omega/\omega' <\infty$~\cite{Lange:2003ff,Grozin:2005iz,Lee:2005gza}.
We also note that we cannot derive the moment-space representation 
of the evolution equation (\ref{evol}) in a usual way
as in the case of the LCDAs for the light mesons~\cite{Braun:1990iv},
%, $\pi$, $\rho$, etc.~\cite{Braun:1990iv},
because
the presence of the logarithm (\ref{L}) prevents us
from performing the Taylor expansion of (\ref{evol}) about $t=0$; indeed,
(\ref{ren1}) shows that the renormalized LCDA is non-analytic at 
$t=0$ (see also the discussion in \cite{Braun:2003wx,Kawamura:2008vq}).
% can be shown that the LCDA (\ref{eq1}) is not Taylor expandable about $t=0$).
Thus, the evolution equation for the $B$-meson LCDA manifests 
simple geometrical structure as the quasilocality of the kernel
only in the coordinate-space representation (\ref{evol}).

One may anticipate that the evolution equation (\ref{evol})
would hold to all orders in perturbation theory
by taking into account the higher-loop terms in the RG functions (\ref{rgf1})-(\ref{rgf2}).
This is correct, at least, for a particular class of higher-loop corrections
associated with the universal cusp anomalous dimension $\Gamma_{\rm cusp}(\alpha_s)$
of Wilson loops.
%in particulatr, 
For example, when we take into account the diagrams that correspond to 
the two-loop corrections to
the relevant Wilson line (\ref{wl}),
$\Gamma_{\rm cusp}(\alpha_s)$ of (\ref{rgf1}) gets modified 
%in (\ref{evol}) 
into~\cite{Korchemsky:1987wg}
\be 
\Gamma_{\rm cusp}(\alpha_s)= \Gamma_{\rm cusp}^{(1)}\frac{\alpha_s}{4\pi}
+\Gamma_{\rm cusp}^{(2)}\left(\frac{\alpha_s}{4\pi}\right)^2\ ,
\label{cusp2}
\ee
with~\cite{Kodaira:1981nh}
\be
%\Gamma_{\rm cusp}^{(1)}=4C_F,~~~~~
\Gamma_{\rm cusp}^{(2)}=4C_F
\left[\left(\frac{67}{9}-\frac{\pi^2}{3}\right)C_G-\frac{10}{9}N_f\right]\ ,
\label{kodaira}
\ee
where $C_G=N_c$ and $N_f$ denotes the number of active flavors. 
Actually, it is not known at present whether the effects of all the other two-loop 
corrections to the bilocal operator in (\ref{eq1})
can be absorbed into the remaining
two RG functions in (\ref{evol}), $\gamma_F(\alpha_s)$ and $K(z, \alpha_s)$,
as their two-loop terms.
Still, the evolution equation (\ref{evol}), with (\ref{rgf12}), (\ref{rgf2}) and 
the two-loop cusp anomalous dimension (\ref{cusp2}) taken into account,
is useful for resumming the Sudakov logarithms to a consistent accuracy,
as we will demonstrate in the next section.

\section{Analytic solution in the coordinate space}
\label{sec3}

The LO solution for the evolution equation of the $B$-meson LCDA was 
obtained in the momentum representation in \cite{Lange:2003ff},
and the result determines $\phi_{+}(\omega, \mu)$ of (\ref{eq1})
as the convolution of $\phi_{+}(\omega', \mu_0)$ at a lower scale $\mu_0$ 
and the (complicated) evolution operator, over the entire range of $\omega'$
(see (\ref{solution:mom}) in Appendix A).
Its Fourier transformation in principle 
% the result obtained in  \cite{Lange:2003ff}.
gives the solution for our evolution equation, (\ref{evol})-(\ref{rgf2}), 
in the coordinate space,
but, in practice, we find it more useful
to solve (\ref{evol}) directly:
mathematically, 
(\ref{evol}) is an integro-differential equation of similar type as 
the corresponding equation in the momentum space
and has simpler structure for the kernel of integral operator 
than the latter case,
as noted in Sec.~\ref{sec2}. 
This would imply that the strategy
devised in \cite{Lange:2003ff} to solve the evolution equation for the latter case
should also allow us to solve (\ref{evol}), possibly with simpler manipulations.
Moreover,
intermediate steps of those manipulations
reveal peculiar structures
behind a rather simple final form of our solution,
% of (\ref{evol}), i.e., 
(\ref{solution:4}) below.
%We solve (\ref{evol})

First of all, we demonstrate that the strategy of \cite{Lange:2003ff} is applicable to 
(\ref{evol}) and allows us to 
construct its general solution
which is exact even when the higher-loop terms in the RG functions
$\Gamma_{\rm cusp} (\alpha_s)$, $\gamma_F(\alpha_s)$,
$K(z, \alpha_s)$ are taken into account.
For this purpose, we further put forward the above-mentioned similarity 
between (\ref{evol}) and the corresponding 
integro-differential equation of \cite{Lange:2003ff} in the momentum space,
%We start our discussion To begin with, we 
by performing the analytic continuation for (\ref{evol}) as $t\rightarrow -i\tau$. 
Then 
the evolution equation (\ref{evol}) becomes the integro-differential equation
for the $B$-meson LCDA at imaginary light-cone separation, $\tilde{\phi}_+(-i\tau,\mu)$,
as
\be
\mu\frac{d}{d\mu} \tilde{\phi}_+(-i\tau,\mu) =
%\biggl\{-\biggl[
-\left[ \Gamma_{\rm cusp}(\alpha_s) \ln(\tau\mu e^{\gamma_E}) +\gamma_F(\alpha_s)
\right] \tilde{\phi}_+(-i\tau,\mu)
%\delta(1-u)
+\int_0^1dz
K(z,\alpha_s)
%\biggr\} 
\tilde{\phi}_+(-iz\tau,\mu)\ .
\label{evol2}
\ee
%can be in principle obtained by Fourier transforming the result obtained in  \cite{Lange:2003ff}.
We recall that the kernel $K(z,\alpha_s)$ corresponds to the coordinate-space
representation 
of a DGLAP-type splitting function and thus can be diagonalized in the moment space.
In the coordinate-space language~\cite{Balitsky:1987bk},
the corresponding moment is given as
\bea
{\cal K}(j, \alpha_s)=\int_0^1dz z^j K(z,\alpha_s)
= {\cal K}^{(1)}(j )\frac{\alpha_s}{4\pi}+  \cdots\ ,
\label{tGam}
\eea
where (\ref{rgf2}) gives the coefficient for the order $\alpha_s$ term as
\bea
{\cal K}^{(1)}(j ) = 4C_F\int_0^1dz z^j
\left(\frac{z}{1-z}\right)_+= -4C_F[\psi(j+2)+\gamma_E-1]\ ,
\label{tGam1}
\eea
with $\psi(z)=(d/dz)\ln \Gamma(z)$ being the di-gamma function, 
and the ellipses in (\ref{tGam}) stand for the (presently unknown) terms 
of order $\alpha_s^2$ and higher.
%playing the role of the anomalous dimension depending on $\lambda$.
As mentioned below (\ref{ft2}), however, 
the usual moment is not useful for treating 
(\ref{evol2}): the presence of logarithm $\ln(\tau\mu e^{\gamma_E})$
in the RHS suggests that 
the 
%(fractional) 
values for the moment $j$ will be modified
%evolve (generally by fractional amounts)
%evolve 
under the variation of the scale $\mu$.
The authors in \cite{Lange:2003ff} demonstrated that
%such type of extention for 
taking into account the corresponding ``evolution'' of the moment $j$ 
indeed enables them to construct
%sufficient for constructing 
the general solution
of the corresponding 
integro-differential equation in the momentum space (see also \cite{Grozin:2005iz}).
Thus, we take the ansatz,
\be
\tilde{\phi}_+(-i\tau,\mu)= \frac{1}{2\pi i} \int_{c-i\infty}^{c+i\infty}dj
\left(\tau \mu_0 e^{\gamma_E}\right)^{j-\xi(\mu,\mu_0)}\varphi (j,\mu)\ ,
\label{ansatz}
\ee
with a real constant $c$,
% is a real constant and we assume $\xi(\mu_0, \mu_0)=0$,
and we determine $\varphi (j,\mu)$ and $\xi(\mu,\mu_0)$ such that (\ref{ansatz}) obeys 
(\ref{evol2}).
This ansatz has the form similar to
the inverse Mellin transformation to construct the solution for
the DGLAP-type evolution equation in the coordinate-space language 
(see \cite{Balitsky:1987bk}), except for the contribution of $\xi(\mu, \mu_0)$,
which describes the evolution of the power of $\tau$
from a certain (low) scale $\mu_0$ to the scale $\mu$.
We assume $\xi(\mu_0, \mu_0)=0$, without loss of generality, 
and $\mu_0$ multiplied by $e^{\gamma_E}$ is put in the integrand of 
(\ref{ansatz}) for convenience.
Substituting (\ref{ansatz}) into (\ref{evol2}), we obtain
\bea
\mu\frac{d}{d\mu} \varphi(j,\mu)&&\!\!\!\!\!\! =
%\biggl\{-
\biggl[
-\Gamma_{\rm cusp}(\alpha_s) \ln(\tau\mu e^{\gamma_E}) -\gamma_F(\alpha_s)
+{\cal K}\left(j-\xi(\mu,\mu_0),\ \alpha_s\right)
\nonumber\\
&&
\left. +\ \mu\frac{d\xi(\mu,\mu_0)}{d\mu} \ln(\tau\mu_0 e^{\gamma_E}) \right] \varphi(j,\mu)\ .
\label{evol3}
\eea
Because the RHS of this equation should be independent of 
%In order for the RHthis equation is independent of 
$\tau$, $\xi(\mu,\mu_0)$ obeys
\be
\mu\frac{d\xi(\mu,\mu_0)}{d\mu} =\Gamma_{\rm cusp}(\alpha_s)\ .
\label{rho}
\ee 
This shows that $\xi(\mu,\mu_0)$ is independent of $j$ and is integrated to give, introducing  
the $\beta$ function, $\beta(\alpha_s)=\mu d\alpha_s / d\mu$,
\be
\xi(\mu, \mu_0) = \int_{\alpha_s(\mu_0)}^{\alpha_s(\mu)}
\frac{\Gamma_{\rm cusp}(\alpha)}{\beta(\alpha)}d\alpha
\equiv \Xi(\alpha_s(\mu),\alpha_s(\mu_0))\ .
\label{func:g}
\ee
Now (\ref{evol3}) reduces to
\be
\mu\frac{d}{d\mu} \varphi(j,\mu) =
%\biggl\{-\biggl[
\left[ -\Gamma_{\rm cusp}(\alpha_s) \ln\frac{\mu}{\mu_0} -\gamma_F(\alpha_s)
+{\cal K}\left( j-\xi(\mu,\mu_0),\ \alpha_s\right)\right] \varphi(j,\mu)\ ,
\label{evol4}
\ee
and this simple differential equation is immediately solved to give
\bea
\varphi(j, \mu)=\exp \left[ {\cal V}(\mu,\mu_0)+ {\cal W}(\mu,\mu_0,j)\right] 
\varphi(j, \mu_0)\ ,
\label{func:h}
\eea
where 
\bea
{\cal V}(\mu,\mu_0)&& \!\!\!\!\!
=-\int_{\alpha_s(\mu_0)}^{\alpha_s(\mu)}\frac{d\alpha}{\beta(\alpha)}
\left[\Gamma_{\rm cusp}(\alpha)\int_{\alpha_s(\mu_0)}^{\alpha}
\frac{d\alpha^\prime}{\beta(\alpha^\prime)}+\gamma_F(\alpha) \right]\ ,
\label{func:V}\\
{\cal W}(\mu,\mu_0,j)&&\!\!\!\!\!=
\int_{\alpha_s(\mu_0)}^{\alpha_s(\mu)}\frac{d\alpha}{\beta(\alpha)}
{\cal K}\left(j -\Xi\left(\alpha, \alpha_s(\mu_0)\right) ,\ \alpha \right)\ ,
\label{func:W}
\eea
and $\varphi(j,\mu_0)$ should be expressed 
%is determined from (\ref{ansatz})
by the Mellin transform of 
the initial condition, $\tilde{\phi}_+(-i\tau,\mu_0)$, from (\ref{ansatz}) (see (\ref{initial})
below).
Substituting these results
%(\ref{func:h})-(\ref{initial}) 
into (\ref{ansatz}),
%and using (\ref{func:g}),
we obtain 
\be
\tilde{\phi}_+(-i\tau,\mu)= 
e^{{\cal V}(\mu,\mu_0)} (\tau \mu_0 e^{\gamma_E})^{-\xi}
\int_0^{\infty}\frac{d\tau^\prime}{\tau^\prime}
\tilde{\phi}_+(-i\tau^\prime,\mu_0)
\int_{c-i\infty}^{c+i\infty}\frac{dj}{2\pi i}
 \left(\frac{\tau}{\tau^\prime}\right)^{j}e^{{\cal W}(\mu,\mu_0,j)}\ .
%h(\tau,\mu.\mu_0;is) \ .
\label{solution:2}
\ee
Here and below, $\xi \equiv  \xi(\mu, \mu_0)$, unless otherwise indicated.
The formula (\ref{solution:2}) in principle gives the solution for (\ref{evol2}),
which is exact even when the higher-order terms in 
$\Gamma_{\rm cusp} (\alpha_s)$, $\gamma_F(\alpha_s)$,
$K(u, \alpha_s)$ are taken into account.
However, this solution has been obtained by assuming tacitly that
$\varphi(j, \mu)$ of (\ref{ansatz}), expressed by the Mellin transform
of $\left(\tau \mu_0 e^{\gamma_E}\right)^{\xi} \tilde{\phi}_+(-i\tau,\mu)$ as
\be
\varphi(j, \mu)=\int_{0}^{\infty}\frac{d\tau}{\tau}
(\tau\mu_0e^{\gamma_E})^{-j} \left(\tau \mu_0 e^{\gamma_E}\right)^{\xi}
\tilde{\phi}_+(-i\tau, \mu)\ ,
\label{initial}
\ee
is a regular function in a certain ``band'' of region in the complex $j$ plane,
and that the constant $c$ in (\ref{solution:2}) is chosen such that
the integration contour in this formula is contained within this band.
Now we consider the condition for the convergence of the integral in (\ref{initial}),
which in turn determines this band,
%the band where $\varphi(j, \mu)$ is regular, 
as well as
the range where (\ref{solution:2}) is applicable:
the short-distance behavior of $\tilde{\phi}_+(-i\tau, \mu)$ as $\tau \rightarrow 0$ in the integrand 
of (\ref{initial}) 
can be determined~\cite{Lee:2005gza,Kawamura:2008vq} by perturbation theory,
%rigorously by the OPE of 
as a constant modulo $\ln\tau$ (see (\ref{ope}) below),
so that (\ref{initial}) is convergent for the integration region $\tau \sim 0$
when $\xi > {\rm Re}(j)$. 
On the other hand, studies of the IR structure of the DA
indicate $\tilde{\phi}_+(-i\tau, \mu)\sim 1/\tau^2$ or more strongly suppressed
as $\tau \rightarrow \infty$~\cite{Grozin:1997pq,KKQT}, so that
the integral in (\ref{initial}) is convergent as $\tau \rightarrow \infty$
when $\xi < {\rm Re}(j)+2$. 
These considerations show that (\ref{initial}) gives a regular function for 
the band with $\xi-2 < {\rm Re}(j) < \xi$ in the complex
$j$ plane, and the constant $c$ in (\ref{ansatz}), (\ref{solution:2}) should be chosen as
\be
\xi-2 < c < \xi\ .
\label{cband}
\ee
Because $\xi$ of (\ref{func:g}) grows from 0,
as $\mu$ increases from $\mu_0$ (see (\ref{rgf1}), (\ref{cusp2}), and (\ref{kodaira})),
only for the values of scales $\mu$ and $\mu_0$ 
satisfying
\be
\xi=\xi(\mu, \mu_0) <2\ ,
\label{xiband}
\ee
$c$ can be chosen as a fixed constant and thus the solution (\ref{solution:2}) describes
the exact evolution of the $B$ meson LCDA from $\mu_0$ to $\mu$.
%\footnote{
(Note that the condition for the convergence of the convolution integrals of
(\ref{solution:2}) and the corresponding hard part in the QCD factorization
formula for exclusive $B$ decays eventually requires (\ref{xiband2}) below.)
%}

To proceed further, 
we change the integration variable in (\ref{func:W}) from $\alpha$ to 
$\Xi(\alpha, \alpha_s(\mu_0))$.
Defining $\alpha_x$ such that $\Xi\left(\alpha_x, \alpha_s(\mu_0)\right)=x$,
we obtain
\be
{\cal W}(\mu,\mu_0,j)
%&&\!\!\!\!\!\!\!
=
\int_{0}^{\xi}dx\frac{{\cal K}\left(j - x,\ \alpha_x \right)}{\Gamma_{\rm cusp}(\alpha_x)}
%&&
= \int_0^{\xi} dx\
\frac{{\cal K}^{(1)}(j-x)}{\Gamma_{\rm cusp}^{(1)}} +\cdots\ ,
\label{func:W2old}
\ee
where the ellipses stand for the NLO or higher contributions 
%associated with the two
that involve the two- or higher-loop anomalous dimensions.
%$\Gamma_{\rm cusp}^{(n)}$ ($n \ge 2$) or ${\cal K}^{(n)}$ ($n \ge 2$).
Using (\ref{rgf1}) and (\ref{tGam1}),
one finds 
\be
e^{{\cal W}(\mu,\mu_0,j)}=e^{\left(1-\gamma_{E}\right) \xi}\
%+\ln  
\frac{\Gamma (j+2 - \xi)}{\Gamma (j+2 )}\ ,
\label{func:W2}
\ee 
up to the corrections 
of the two-loop level.
In the complex $j$ plane, (\ref{func:W2}) has poles at
$j= \xi -2- n$ with $n=0,1,\cdots$,
which 
%Because these poles 
are all located in the left of the integration contour 
in (\ref{solution:2})
with (\ref{cband}):
by the theorem of residues,
these poles give rise to nonzero contribution to the $j$ integral 
%in (\ref{solution:2})
for $\tau>\tau^\prime$, while the $j$ integral vanishes for $\tau<\tau^\prime$.
Evaluation of those pole contributions yields
\bea
\int_{c-i\infty}^{c+i\infty}\frac{dj}{2\pi i}
 \left(\frac{\tau}{\tau^\prime}\right)^{j}\frac{\Gamma (j+2 - \xi)}{\Gamma (j+2 )}
%e^{{\cal W}(\mu,\mu_0,j)}
&& \!\!\!\!\!\!
=\theta(\tau-\tau^\prime)\sum_{n=0}^{\infty}
\frac{(-1)^n}{n!\Gamma( \xi- n)}\left(\frac{\tau}{\tau^\prime}\right)^{\xi-2- n}
\nonumber\\
&&\!\!\!\!\!\!
=\theta(\tau-\tau^\prime) \frac{\left( \tau^\prime/\tau\right)^{2- \xi}}
{\left(1-\tau^\prime/\tau\right)^{1- \xi}\Gamma( \xi)}\ .
\label{poles}
\eea
Substituting this result into (\ref{solution:2}) and changing the integration variable from
$\tau^\prime$ to $z=\tau^\prime/\tau$, we obtain
\bea
\tilde{\phi}_+(-i\tau,\mu)= e^{{\cal V}(\mu,\mu_0)} \left(\tau \mu_0e^{\gamma_E}\right)^{-\xi}
%\times
\frac{e^{(1-\gamma_E)\xi}}{\Gamma(\xi)}
\int_0^1dz  \left(\frac{z}{1-z}\right)^{1-\xi} \tilde{\phi}_+(-i\tau z,\mu_0)\ ,
\label{solution:4}
\eea
which is exact up to the NLO corrections mentioned in (\ref{func:W2old}).
The contribution of the kernel (\ref{rgf2}) 
%at the LO 
in perturbation theory
receives the RG improvement in (\ref{solution:4}) as $(z/[1-z])^{1-\xi}$ 
with the modified power $1-\xi$, 
where $\xi$ of (\ref{func:g}) is
induced by the 
%one-loop 
cusp anomalous dimension.
%(see (\ref{func:g})).
For the case with $\xi \rightarrow 0$, we have
the singular behavior 
%in the integrand 
as
%the integrand has the singularity at $z = 1$ as
$1/(1-z)^{1-\xi} =(1/\xi)\delta (1-z) +1/(1-z)_+ +O(\xi)$,
%the integrand, and 
but this eventually gives the finite contribution to the RHS of (\ref{solution:4}),
combined with the $\xi \rightarrow 0$ behavior of the gamma function, 
$\Gamma (\xi)=1/\xi-\gamma_E +O(\xi)$.
This also shows that the RHS of (\ref{solution:4}) reduces to
$\tilde{\phi}_+(-i\tau,\mu_0)$ when $\mu \rightarrow \mu_0$, i.e., when
% and thus
$\xi \rightarrow 0$
and ${\cal V}(\mu, \mu_0) \rightarrow 0$
(see (\ref{func:g}), (\ref{func:V})), as it should be.
In (\ref{solution:4}),
% with (\ref{U}), 
it is straightforward to 
perform the analytic continuation from 
the imaginary light-cone separation to the real one, as $\tau \rightarrow it$,
and the resulting solution for the evolution equation (\ref{evol})
embodies a quite simple structure 
to determine the $B$-meson LCDA 
with a quark-antiquark light-cone 
separation $t$ 
in terms of the LCDA at a lower renormalization scale $\mu_0$ 
with
smaller 
interquark 
separations.
The Fourier transformation of this result is
calculated 
in Appendix~\ref{appa},
and the obtained momentum representation
(\ref{solution:mom})
reproduces the result of \cite{Lange:2003ff,Lee:2005gza}
derived 
in the momentum space;
in particular, 
the factor $(\tau\mu_0 e^{\gamma_E})^{-\xi}$ in (\ref{solution:4}), 
which is non-analytic at $\tau \rightarrow 0$,
produces the 
%hard 
radiative tail 
as $\sim \omega^{\xi-1}$ for large $\omega$ in (\ref{solution:mom}),
which renders 
%the $\tau\rightarrow 0$ limit of $\tilde{\phi}_+(\tau,\mu)$ and 
%its derivatives singular, which corresponds to the divergence of 
all non-negative moments of the LCDA, 
$\int_0^\infty d\omega\omega^n \phi_+(\omega,\mu)$ with $n=0,1,2, \ldots$, 
divergent,
irrespective of the initial behavior, $\phi_+(\omega,\mu_0)$~\cite{Lange:2003ff}.
%\bea
%\left.\frac{\der^n}{\der\tau^n}\right|_{\tau=0}\tilde{\phi}_+(-i\tau,\mu)
%=\int_0^\infty d\omega\omega^n \phi_+(\omega,\mu) \ .
%\eea 
We also emphasize that our result (\ref{solution:4}) 
has a much simpler structure than (\ref{solution:mom});
i.e., the most compact expression possible for calculating the 
%one-loop 
evolution of the $B$-meson LCDA under changes of the renormalization scale
is provided by our coordinate-space result (\ref{solution:4}).

We note that
the first two factors
in (\ref{solution:4}), given by
\be
e^{{\cal V}(\mu,\mu_0)} \left(\tau \mu_0e^{\gamma_E}\right)^{-\xi}
=e^{{\cal V}(\mu,\mu_0)-\xi \ln \left(\tau \mu_0e^{\gamma_E}\right)}\ ,
\label{prefactor}
\ee
are unaffected by the above 
manipulations (\ref{func:W2}), (\ref{poles}),
which are valid up to the NLO corrections,
and thus (\ref{prefactor}) gives the exact result
even when 
the higher-order terms in 
$\Gamma_{\rm cusp} (\alpha_s)$ and $\gamma_F(\alpha_s)$
are taken into account for (\ref{func:g}) and (\ref{func:V}).
Indeed, substituting those definitions of $\xi$ and ${\cal V}(\mu,\mu_0)$,
we may reexpress the exponent of the RHS in (\ref{prefactor}) as
\be
{\cal V}(\mu,\mu_0)-\xi \ln \left(\tau \mu_0e^{\gamma_E}\right)
= -\int_{\mu_0}^{\mu}\frac{d\mu'}{\mu'}\left[\Gamma_{\rm cusp}(\alpha_s(\mu'))
\ln\left(\tau \mu' e^{\gamma_E}\right) +\gamma_F(\alpha_s(\mu')) \right]\ ,
\label{U2}
\ee
%using (\ref{func:g}), (\ref{func:V}).
which shows that (\ref{prefactor}) 
corresponds to
the general solution of the evolution equation (\ref{evol2}) with 
%$K(z, \alpha_s)= 0$ substituted.
the contribution of the kernel $K(z, \alpha_s)$ omitted~\cite{foot3}.
%\footnote{
%The RG evolution of the shape function for inclusive
%$B$ decays is described by the evolution operator of this type, see \cite{KS94}.} 
We now derive
the explicit form of (\ref{func:g}) and (\ref{func:V})
arising in our solution (\ref{solution:2}),  (\ref{solution:4}):
substituting (\ref{rgf1}), (\ref{rgf12}), (\ref{cusp2})
and the usual perturbative expansion 
for the 
$\beta$ function,
\bea
\beta(\alpha_s)&& \!\!\!\!\!\!
=\mu\frac{d\alpha_s}{d\mu} 
= -2 \alpha_s \sum_{n=0}^\infty \beta_n \left(\frac{\alpha_s}{4\pi}\right)^{n+1}\ ,
\nonumber\\
\beta_0&& \!\!\!\!\!
=\frac{11}{3}C_G-\frac{2}{3}N_f\ ,~~~~
\beta_1=\frac{34}{3}C_G^2-\frac{10}{3}C_GN_f-2C_FN_f\ ,~
\cdots\ ,
\eea
a straightforward calculation gives 
\be
\xi(\mu,\mu_0) = \frac{\Gamma_{\rm cusp}^{(1)}}{2\beta_0}
\left\{ 
\ln\frac{\alpha_s(\mu_0)}{\alpha_s(\mu)}
+\frac{\alpha_s(\mu_0)-\alpha_s(\mu)}{4\pi}
\left(\frac{\Gamma_{\rm cusp}^{(2)}}{\Gamma_{\rm cusp}^{(1)}}
-\frac{\beta_1}{\beta_0}\right) 
\right\}+ \cdots\ ,
\label{func:g1}
\ee
and
\bea
\lefteqn{{\cal V}(\mu,\mu_0)
= \frac{\Gamma_{\rm cusp}^{(1)}}{4 \beta_0^2}
\left\{\frac{4\pi}{\alpha_s(\mu_0)}\left(1+ \ln\frac{\alpha_s(\mu_0)}{\alpha_s(\mu)}\right)
- \frac{4\pi}{\alpha_s(\mu)}\right\}
-\frac{\gamma_F^{(1)}}{2\beta_0}\ln\frac{\alpha_s(\mu_0)}{\alpha_s(\mu)}}  
\nonumber\\
&& +\frac{\Gamma_{\rm cusp}^{(1)}}{4 \beta_0^2}
\left\{
\frac{\beta_1}{2\beta_0}\ln^2 \frac{\alpha_s(\mu_0)}{\alpha_s(\mu)}
+ \left(\frac{\Gamma_{\rm cusp}^{(2)}}{\Gamma_{\rm cusp}^{(1)}}
-\frac{\beta_1}{\beta_0}\right)
\left(\frac{\alpha_s(\mu_0)- \alpha_s(\mu)}{\alpha_s(\mu_0)}- 
\ln\frac{\alpha_s(\mu_0)}{\alpha_s(\mu)}\right)
\right\} +\cdots\ ,
\label{func:V1}
\eea
where the ellipses stand for the terms 
that are down by $\alpha_s$ compared with the preceding terms 
and receive the contributions due to higher loops, e.g., 
the three-loop cusp anomalous dimension $\Gamma_{\rm cusp}^{(3)}$, 
the two-loop local anomalous dimension
%or with the 
$\gamma_F^{(2)}$, etc.
If we substitute only the one-loop terms of these results,
the first term 
of (\ref{func:g1}) and the first line of (\ref{func:V1}),
into (\ref{solution:4}),
we obtain the explicit analytic form of the solution 
for the evolution equation (\ref{evol}),
exact at the one-loop level with (\ref{rgf1})-(\ref{rgf2}).

The definition (\ref{func:V}) shows that 
${\cal V}(\mu, \mu_0)$ involves the contribution 
associated with the first term in the RHS of (\ref{evol4}), i.e.,
the cusp anomalous dimension accompanying $\ln(\mu/\mu_0)\sim 1/\alpha_s$.
As a result, in (\ref{func:V1}),
the contributions associated with the cusp anomalous dimension 
are enhanced by the factor induced by this logarithm, compared to the contribution from
the second term of (\ref{func:V}) with the local anomalous dimension (\ref{rgf12}):
the leading term is given by the one-loop cusp anomalous dimension $\Gamma_{\rm cusp}^{(1)}$,
while the one-loop local anomalous dimension $\gamma_F^{(1)}$ contributes to the 
next-to-leading term, i.e., at the same level
as the two-loop 
cusp anomalous dimension.
Here, the contribution due to $\gamma_F^{(1)}$ corresponds to the one-loop 
level in
the usual RG-improved perturbation theory, and thus the treatment at this level 
has to be complemented with the two-loop contributions associated with 
the cusp anomalous dimension,
the second line of (\ref{func:V1}).
This pattern is characteristic of the Sudakov-type large logarithmic effects
induced by the cusp anomalous dimension.
This fact also 
requires us to reorganize our result (\ref{solution:4}) as well as (\ref{func:V1})
according to consistent order counting 
of those logarithmic contributions.
This can be achieved 
by introducing
\be
\chi = \beta_0\frac{\alpha_s(\mu)}{4\pi}\ln\frac{\mu^2}{\mu_0^2}\ ,
\label{lambda}
\ee
and by following the standard procedure used 
in the soft gluon resummation formalism in QCD~\cite{Catani:2003zt}:
we organize (\ref{func:V1})
by a systematic large logarithmic expansion,
where $\chi$ is formally considered of order unity and the small expansion parameter 
is $\alpha_s(\mu)$,
leading to
\be
\ln\frac{\alpha_s(\mu_0)}{\alpha_s(\mu)}
=-\ln(1-\chi)-\frac{\alpha_s(\mu)}{4\pi}\frac{\beta_1}{\beta_0}\frac{\ln(1-\chi)}{1-\chi}
+O(\alpha_s^2)\ .
\label{lna}
\ee
Substituting this expansion, 
(\ref{func:V1}) is recast into 
\bea
{\cal V}(\mu,\mu_0)
=
\frac{4\pi}{\alpha_s(\mu)}h^{(0)}(\chi)+h^{(1)}(\chi)\ , 
\label{exponent}
\eea
up to the corrections of $O(\alpha_s)$, with
\bea
\!\!\!\!\!\! h^{(0)}(\chi)&&\!\!\!\!\!=\frac{\Gamma_{\rm cusp}^{(1)}}{4\beta_0^2}
\left[\left(\chi-1\right)\ln(1-\chi)
-\chi\right] \ ,
\label{func:h0}
\\
\!\!\!\!\!\! h^{(1)}(\chi)&&\!\!\!\!\!=\frac{\Gamma_{\rm cusp}^{(1)}}{4\beta_0^2}
\left[-\frac{\beta_1}{2\beta_0}\ln^2(1-\chi)
+\left(\frac{\Gamma_{\rm cusp}^{(2)}}{\Gamma_{\rm cusp}^{(1)}}
-\frac{\beta_1}{\beta_0}\right)\left(\ln(1-\chi)+\chi\right)\right]
+
\frac{\gamma_F^{(1)}}{2\beta_0}
\ln(1-\chi) \ .
\label{func:h1}
\eea
In (\ref{exponent}),
the first and the second terms, $(4\pi/\alpha_s)h^{(0)}(\chi)$ and $h^{(1)}(\chi)$, 
collect the terms
$\alpha_s^n\ln^{n+1}(\mu^2/\mu_0^2)$ and $\alpha_s^n \ln^{n}(\mu^2/\mu_0^2)$, respectively,  
with $n=1,2, \ldots$,
corresponding to the LL and NLL contributions,
while the $O(\alpha_s)$ corrections omitted from (\ref{exponent}) 
correspond to the NNLL or higher level.
Thus, the first factor $e^{{\cal V}(\mu,\mu_0)}$
in (\ref{solution:4})
is the exponentiation of the logarithmic terms $\alpha_s^n \ln^{m}(\mu^2/\mu_0^2)$
with $m\le n+1$, 
playing analogous role 
as the Sudakov form factor in the soft gluon resummation in QCD~\cite{Catani:2003zt}.
It is straightforward to see that this factor $e^{{\cal V}(\mu,\mu_0)}$
with only the LL term, 
$(4\pi/\alpha_s)h^{(0)}(\chi) 
= -(\alpha_s/32\pi)\Gamma_{\rm cusp}^{(1)}\ln^2(\mu^2/\mu_0^2)+ \cdots$,
retained in the exponent (\ref{exponent}) 
corresponds to the double leading logarithmic approximation
summing up the towers of logarithms $\alpha_s^n \ln^{2n}(\mu^2/\mu_0^2)$,
and $e^{{\cal V}(\mu,\mu_0)}$ with the exponent (\ref{exponent}) 
at the NLL accuracy
resums 
the first three towers of logarithms,
$\alpha_s^n \ln^{m}(\mu^2/\mu_0^2)$ with $m=2n$, $2n-1$, and $2n-2$, 
exactly,
to all orders in $\alpha_s$.

The logarithmic expansion (\ref{lna}) can be also applied to (\ref{func:g1}), 
yielding
\be
\xi(\mu, \mu_0)= - \frac{\Gamma_{\rm cusp}^{(1)}}{2\beta_0}\ln(1-\chi)\ ,
\label{func:g3}
\ee
up to the corrections of $O(\alpha_s)$.
This result does not receive the logarithmic enhancement
but
obeys order counting similar as the contribution from 
the second term in (\ref{func:V}) due to the local anomalous dimension $\gamma_F$,
%because 
as apparent comparing (\ref{func:g}) with (\ref{func:V}).
Thus, the substitution of (\ref{func:g3}) into (\ref{solution:4})
produces
the NLL-level contributions
while the omitted $O(\alpha_s)$ contributions correspond to the NNLL or higher level,
using the order counting similar as in (\ref{exponent}).
Therefore, our solution~(\ref{solution:4}) with (\ref{exponent})-(\ref{func:g3})
substituted embodies the evolution of the $B$-meson LCDA,
accomplishing the relevant Sudakov resummation, and 
is exact up to the corrections of the NNLL-level. 
We note that controlling
the NNLL-level effects completely
%for the exponent (\ref{exponent})
requires to take into account 
the three-loop cusp anomalous dimension $\Gamma^{(3)}_{\rm cusp}$,
as well as the local anomalous dimension and DGLAP-type splitting function at the two-loop level,
$\gamma_F^{(2)}$ and ${\cal K}^{(2)}(j)$, in (\ref{solution:2}) with 
(\ref{func:V}), (\ref{func:W}).

Before ending this section,
we mention
that the factor $\ln(\tau\mu_0 e^{\gamma_E})$ accompanying $\xi$ 
in the RHS of 
(\ref{prefactor})
could produce
an additional logarithmic 
enhancement. 
Also, 
%could arise 
in the integrand of 
(\ref{solution:4}), 
%i.e., the effect of $g^{(2)}(\lambda)$, combined with 
the behavior 
as $z\rightarrow 0$ and $z\rightarrow 1$ could receive 
another logarithmic enhancement.
These facts suggest that
%and thus 
the higher-order terms associated with the similar types of logarithms could be relevant 
if we intend to determine the precise 
shape of the LCDA 
at the ``edge''.
However, we do not go into the details of such higher-order effects here:
systematic treatment of those higher-order effects 
would require
an approach, which is beyond the scope of this work
based on the evolution equation for the renormalization scale $\mu$.
Furthermore, it is the integrals of the LCDA over $\tau$, like 
(\ref{lnmom}) below,
that is eventually relevant to exclusive $B$ decays, and the above types of 
higher-order logarithmic effects
at the edge region should play minor roles on the value of those convergent integrals,
where the only relevant logarithm is $\ln (\mu^2/\mu_0^2)$ treated in this section.

\section{Master formula for the inverse moments of the LCDA}
\label{sec4}

The $B$-meson LCDA (\ref{eq1})
with $\mu = \mu_{\rm hc}$ participates in 
the QCD factorization formula for exclusive 
$B$ decays~\cite{Beneke:2000ry,Beneke:2000wa,Beneke:2005vv,Bell08}
through the inverse moments, 
\be
\lambda_B^{-1}(\mu)
\equiv\int_0^\infty \frac{d\omega}{\omega} \phi_+(\omega,\mu)\ ,
\;\;\;\;\;\;\;\;\;\;
\sigma_n(\mu) 
\equiv
\lambda_B(\mu) \int_0^\infty 
\frac{d\omega}{\omega}\phi_+(\omega,\mu) \ln^n \frac{\mu}{\omega}\ .
\label{sigman}
\ee
Here, $\lambda_B^{-1}(\mu)$
appears as the convolution with the hard part 
in the LO for the hard spectator amplitudes, 
while the calculation of the NLO effects for the hard spectator amplitudes 
requires also the logarithmic moments $\sigma_n(\mu)$ with $n=1, 2$. 
We introduce the logarithmic moments in the coordinate space,
\be
R_n(\mu) \equiv \int_0^\infty  d \tau \tilde{\phi}_+  ( - i\tau ,\mu ) \ln^n(\tau\mu e^{\gamma_E})\ ,
\label{lnmom}
\ee
which are related to (\ref{sigman}) as
\be
\lambda^{-1}_B(\mu)=R_0(\mu)\ , \;\;\;\;
\sigma_1(\mu)=\lambda_B(\mu)R_1(\mu)\ , \;\;\;\;
\sigma_2(\mu)=\lambda_B(\mu) R_2(\mu)-\frac{\pi^2}{6}\ , \;\;
\cdots\ .
\label{lsr}
\ee
These relations
can be obtained, e.g., 
by considering the generating function for the inverse moments (\ref{sigman}),
\be
\int_0^\infty \frac{d\omega}{\omega} \left(\frac{\mu}{\omega}\right)^s \phi_+(\omega,\mu)
=\sum_{n=0}^\infty 
\frac{\sigma_n(\mu)}{\lambda_B(\mu)}
\frac{s^n}{n!}\ ,
\label{generate}
\ee
and relating this to the generating function for $R_n(\mu)$ of (\ref{lnmom}),
i.e., $\int_0^\infty  d \tau  (\tau\mu e^{\gamma_E})^s \tilde{\phi}_+  ( - i\tau ,\mu )$, as
\be
\int_0^\infty \frac{d\omega}{\omega} \left(\frac{\mu}{\omega}\right)^s \phi_+(\omega,\mu)
= \frac{e^{-s\gamma_E}}{\Gamma(1+ s)}
\int_0^\infty  d \tau  (\tau\mu e^{\gamma_E})^s \tilde{\phi}_+  ( - i\tau ,\mu )\ ,
\label{generate2}
\ee
where 
$e^{-s\gamma_E}/\Gamma(1+ s)=\exp[-\sum_{k=2}^\infty (-s)^k \zeta(k)/k ]$,
with $\zeta(k)$ being the Riemann zeta-function.
Remarkably, the simple form of our solution~(\ref{solution:4}) allows us to express
the generating function in the RHS of (\ref{generate2}) as
\bea
\int_0^\infty  d \tau  (\tau\mu e^{\gamma_E})^s \tilde{\phi}_+  ( - i\tau ,\mu )
&&\!\!\!\!\!
=e^{{\cal V}(\mu,\mu_0)+(1-\gamma_E)\xi} 
\frac{\Gamma(1-s)}{\Gamma(1-s+\xi)}
\nonumber\\
&&\;\times\left(\frac{\mu}{\mu_0}\right)^s
\int_0^\infty d\tau (\tau\mu_0 e^{\gamma_E})^{s-\xi} \tilde{\phi}_+(-i\tau,\mu_0)\ ,
\label{integral}
\eea
which plays role of the master formula to derive all the relevant moments
at $\mu = \mu_{\rm hc}$ in terms of the integrals of the 
LCDA at a lower scale $\mu_0$:
Taylor expanding the both sides of this formula about $s=0$, 
we immediately find
\be
\lambda_B^{-1}(\mu)
=R_0(\mu)=
\frac{e^{{\cal V}(\mu,\mu_0)+(1-\gamma_E)\xi}}{\Gamma(1+ \xi)}
\int_0^\infty d\tau(\tau\mu_0 e^{\gamma_E})^{- \xi} \tilde{\phi}_+(-i\tau,\mu_0)\ ,
\label{lambda_B}
\ee
and 
\bea
R_1(\mu)&& \!\!\!\!\!=
\frac{e^{{\cal V}(\mu,\mu_0)+(1-\gamma_E)\xi}}{\Gamma(1+ \xi)}
\int_0^\infty d\tau \left[\ln (\tau \mu e^{2\gamma_E}) + \psi( 1 + \xi ) \right]
(\tau\mu_0 e^{\gamma_E})^{- \xi} \tilde{\phi}_+(-i\tau,\mu_0)\ ,
\label{R1}\\
R_2(\mu)&& \!\!\!\!\!=
\frac{e^{{\cal V}(\mu,\mu_0)+(1-\gamma_E)\xi}}{\Gamma(1+ \xi)}
\nonumber\\
&& \!\!\!\!\!\!\!\!
\times
\int_0^\infty d\tau\left\{  \left[ \ln (\tau\mu e^{2\gamma_E}) + \psi( 1 + \xi )\right]^2
- \psi'( 1 + \xi )
+ \frac{\pi^2}{6} \right\}
(\tau\mu_0 e^{\gamma_E})^{- \xi} \tilde{\phi}_+(-i\tau,\mu_0) ,
\label{R2}
\eea
where $\psi'(z) =(d/dz)\psi(z)$; it is also possible to derive the similar formulae
for $R_n(\mu)$ ($n \ge 3$).
These formulae (\ref{lambda_B})-(\ref{R2}), combined with (\ref{lsr}), 
allow us to calculate (\ref{sigman}) 
with 
the LCDA $\tilde{\phi}_+(-i\tau,\mu_0)$
as the input at the hadronic scale,
and the results are exact up to the NNLL corrections 
when substituting (\ref{exponent})-(\ref{func:g3}).
%in (\ref{lambda_B})-(\ref{R2}).
Furthermore, 
it is worth presenting 
%we can transform them
the corresponding results transformed into 
the momentum representation:
substituting (\ref{integral}) into (\ref{generate2}),
we obtain our master formula in the momentum representation as
\bea
\int_0^\infty \frac{d\omega}{\omega} \left(\frac{\mu}{\omega}\right)^s \phi_+(\omega,\mu)
&&\!\!\!\!\!
=e^{{\cal V}(\mu,\mu_0)+(1-2 \gamma_E)\xi} 
\frac{\Gamma(1-s) \Gamma(1+s-\xi)}{\Gamma(1+ s)  \Gamma(1-s+\xi)}
\nonumber\\
&&\;\times\left(\frac{\mu}{\mu_0}\right)^s
%T(s, \mu_0)\ ,
\int_0^\infty \frac{d\omega}{\omega} \left(\frac{\mu_0}{\omega} 
\right)^{s-\xi} \phi_+(\omega,\mu_0)\ ,
\label{integral3}
\eea
and, using (\ref{generate}), we obtain
\bea
\lambda_B^{-1}(\mu)
&& \!\!\!\!\!
=e^{{\cal V}(\mu,\mu_0)+(1-2 \gamma_E)\xi} 
\frac{\Gamma(1-\xi)}{\Gamma(1+\xi)}
\int_0^\infty \frac{d\omega}{\omega} \left(\frac{\omega}{\mu_0}
\right)^{\xi} \phi_+(\omega,\mu_0)\ ,
\label{lambda_B2}\\
\frac{\sigma_1(\mu)}{\lambda_B(\mu)} && \!\!\!\!\!
=e^{{\cal V}(\mu,\mu_0)+(1-2 \gamma_E)\xi} 
\frac{\Gamma(1-\xi)}{\Gamma(1+\xi)}
\nonumber\\
&& \!\!\!\!\!\!
\times
\int_0^\infty \frac{d\omega}{\omega}
\left[\ln \frac{\mu e^{2\gamma_E}}{\omega} + \psi(1-\xi)  + \psi(1+\xi)\right]
\left(\frac{\omega}{\mu_0}
\right)^{\xi} \phi_+(\omega,\mu_0)\ ,
\label{sigma1}\\
\frac{\sigma_2(\mu)}{\lambda_B(\mu)} && \!\!\!\!\!
=e^{{\cal V}(\mu,\mu_0)+(1-2 \gamma_E)\xi} 
\frac{\Gamma(1-\xi)}{\Gamma(1+\xi)}\int_0^\infty \frac{d\omega}{\omega}
\nonumber\\
&& \!\!\!\!\!\!\!\!\!
\times
\left\{ \left[\ln \frac{\mu e^{2\gamma_E}}{\omega} + \psi(1-\xi)  + \psi(1+\xi)\right]^2
+ \psi'(1-\xi)- \psi'(1+\xi)\right\}
\left(\frac{\omega}{\mu_0}
\right)^{\xi} \phi_+(\omega,\mu_0)\ ,
\label{sigma2}
\eea
and so on. We note that 
(\ref{lambda_B2}) and (\ref{sigma1}) reproduce the corresponding results
that were found
in~\cite{Bell:2008er} in the context of the RG evolution at the one-loop level
in the momentum representation,
while the closed form (\ref{sigma2}) for $\sigma_2$, as well as 
the above formulae (\ref{lambda_B})-(\ref{R2}) in the coordinate-space representation,
is new. We also emphasize that our master formulae (\ref{integral}), (\ref{integral3})
allow us to derive the closed form for the higher 
logarithmic moments straightforwardly. 
%which is again simpler than the corresponding expression in momentum space~\cite{Bell:2008er}.

The above results (\ref{integral})-(\ref{sigma2}) 
demonstrate that, as a result of the evolution, 
the relevant logarithmic-moment integrals have the common, 
additional ``weight functions'' determined by $\xi$,
i.e.,  
%additional weight , 
$(\tau \mu_0 e^{\gamma_E})^{-\xi}$ and $(\omega/\mu_0)^\xi$ in the 
coordinate and momentum representations, respectively,
%in their integrand,
compared with the corresponding formulae 
for $\mu = \mu_0$.
In particular, 
the condition for the convergence of the integrals in (\ref{lambda_B})-(\ref{R2}),
taking into account the $\tau \rightarrow 0$ (and $\tau \rightarrow \infty$)
behavior of $\tilde{\phi}_+(-i\tau,\mu_0)$
%which was 
mentioned above (\ref{cband}),
indicates that 
only for the values of scales $\mu$ and $\mu_0$ 
satisfying
\be
\xi=\xi(\mu, \mu_0) <1\ ,
\label{xiband2}
\ee
our formulae (\ref{lambda_B})-(\ref{R2}) (and
(\ref{lambda_B2})-(\ref{sigma2})) are well-defined and applicable.
This is 
indeed 
satisfied for the relevant scales,
%all reasonable values of scales, 
$\mu= \mu_{\rm hc}$ ($\sim \sqrt{m_b\Lambda_{\rm QCD}}$) and
$\mu_0 \sim 1$~GeV,  
as discussed in the applications in the next section (see Fig.~\ref{fig:4} below), 
and, actually, even for quite large values of $\mu$.

The perturbative expansion of the above results (\ref{lambda_B})-(\ref{R2}) 
(or (\ref{lambda_B2})-(\ref{sigma2})) in terms of $\alpha_s(\mu)$ ($\equiv \alpha_s$)
%$(\equiv \alpha_s)$to the leading-order $\alpha_s$ 
yields (see (\ref{lambda}), (\ref{exponent})-(\ref{func:g3}))
\bea
\lambda_B^{-1}(\mu)&&
\!\!\!\!\!\! =
%\approx
\lambda^{-1}_B(\mu_0)
\left[1-\frac{\alpha_sC_F}{2\pi}\ln^2\frac{\mu}{\mu_0}
+\frac{\alpha_sC_F}{2\pi}\ln\frac{\mu}{\mu_0}
\right]-R_1(\mu_0)
\frac{\alpha_sC_F}{\pi}
\ln\frac{\mu}{\mu_0}\ ,
\label{lambda_B:LO}
\\
R_1(\mu)&&\!\!\!\!\!\! =
\lambda^{-1}_B(\mu_0)
\left[ 1 - \frac{\alpha_s C_F}{2\pi}\ln^2 \frac{\mu}{\mu_0} + 
  \frac{\alpha_sC_F}{2\pi} \ln \frac{\mu}{\mu_0} + \frac{\pi \alpha_sC_F}{6} 
\right]\ln\frac{\mu}{\mu_0}
\nonumber\\
&&\!\!\!\!\!\! 
+ R_1(\mu_0) \left[1 - 
     \frac{3\alpha_sC_F}{2\pi}\ln^2 \frac{\mu}{\mu_0}+
      \frac{\alpha_sC_F}{2\pi}\ln \frac{\mu}{\mu_0} 
     \right]
- R_2(\mu_0) \frac{\alpha_sC_F}{\pi}\ln \frac{\mu}{\mu_0}\ ,
\label{R1:LO}
\\
R_2(\mu)&&\!\!\!\!\!\! =
\lambda^{-1}_B(\mu_0)\left[
  \ln \frac{\mu}{\mu_0} - 
  \frac{\alpha_sC_F}{2\pi}\ln^3 \frac{\mu}{\mu_0} + \frac{\alpha_sC_F}{2\pi}\ln^2 \frac{\mu}{\mu_0} 
 + \frac{\pi\alpha_sC_F}{3} \ln \frac{\mu}{\mu_0} 
+ \frac{2\alpha_s C_F}{\pi} \zeta(3) 
\right] \ln \frac{\mu}{\mu_0} 
\nonumber\\ 
&&\!\!\!\!\!\! + R_1(\mu_0)
   \left[ 2 - 
     \frac{2\alpha_s C_F}{\pi}\ln^2 \frac{\mu}{\mu_0} + \frac{\alpha_sC_F}{\pi}\ln \frac{\mu}{\mu_0} 
 + \frac{\pi \alpha_sC_F}{3}
     \right]\ln \frac{\mu}{\mu_0} 
\nonumber\\
&&\!\!\!\!\!\! 
+ R_2(\mu_0)
   \left[ 1- 
     \frac{5\alpha_s C_F}{2\pi}\ln^2 \frac{\mu}{\mu_0} 
+ \frac{\alpha_s C_F}{2\pi}\ln \frac{\mu}{\mu_0} 
     \right]  - R_3(\mu_0) \frac{\alpha_s C_F}{\pi}\ln \frac{\mu}{\mu_0}\ ,
\label{R2:LO}
\eea
etc., to the corrections of order $\alpha_s^2$;
here we have substituted (\ref{rgf1}), (\ref{rgf12})
for the relevant anomalous dimensions.
These relations 
%(\ref{lambda_B:LO})-(\ref{R2:LO}) 
can also be obtained by substituting the expansion (\ref{LO}), discussed below,
into the RHS of (\ref{lnmom}).
Note that 
%the dominant effects of 
the correction terms
in (\ref{lambda_B:LO})-(\ref{R2:LO})
have the relative size $\lesssim (\alpha_s C_F/\pi)\ln^2 (\mu/\mu_0)$,
% \sim \xi$,
compared to the leading terms.
In particular, 
our first relation (\ref{lambda_B:LO}) has the 
%additional 
double logarithmic term,
$\lambda^{-1}_B(\mu_0)[-(\alpha_sC_F/2\pi)\ln^2(\mu/\mu_0)]$,
%compared with 
which was absent in the corresponding relation discussed in 
%eq.(42) in Ref.~
\cite{Braun:2003wx};
this term comes from the expansion of the LL term
of ${\cal V}(\mu, \mu_0)$ (see (\ref{exponent}), (\ref{func:h0})) 
and it is straightforward to check that, taking into account this double logarithmic term,
(\ref{lambda_B:LO}) satisfies the correct evolution equation for $\lambda^{-1}_B(\mu)$,
which is obtained by
integrating (\ref{evol2}) over $\tau$ (see also (\ref{LO}) below).
%As illustrated in those formulae, however, 
The formulae (\ref{lambda_B:LO})-(\ref{R2:LO}) 
allow us to relate the relevant logarithmic moments at $\mu = \mu_{\rm hc}$
(see (\ref{sigman})-(\ref{lsr}))
to the similar moments at $\mu_0 \sim 1$~GeV.
As we demonstrate in the next section,
the formulae (\ref{lambda_B:LO})-(\ref{R2:LO}) actually show good accuracy at the relevant scales
% in a model-independent way,
and thus provide model-independent relations that are expected to be useful 
for the phenomenological applications.
The pattern illustrated in those formulae 
%(\ref{lambda_B:LO})-(\ref{R2:LO}) 
is general, such that
the calculation of $R_n(\mu)$ at the fixed-order $\alpha_s$
requires the knowledge of the $n+2$ logarithmic moments at $\mu_0$, 
$R_k(\mu_0)$ ($k \le n+1$).

\section{Evolution of the OPE-based LCDA}
\label{sec5}

In this section we apply the evolution represented by our solution (\ref{solution:4}) 
to a suitable input LCDA 
at the initial scale $\mu_0$,
and study the behavior 
of the resultant LCDA $\tilde{\phi}_+(t, \mu)$ at a higher scale $\mu$
to clarify the effects of the NLL-level evolution quantitatively.
For the input LCDA,
we note that 
the model-independent information
is now available based on the OPE for the bilocal operator in (\ref{eq1})
as the short-distance expansion 
for the quark-antiquark light-cone separation $t$~\cite{Kawamura:2008vq}, i.e.,
\be
\bar{q}(tn)
%{\rm P}e^{ig\int_0^td\lambda n\cdot A(\lambda n)}
[tn,0]
\Slash{n}
\gamma_5h_v(0) 
\sim \sum_{i}C_i(t,\mu) {\cal O}_i(\mu)\ ,
\label{ope0}
\ee
as $t \rightarrow 0$,
with the local composite operators ${\cal O}_i(\mu)$ 
and the corresponding Wilson coefficients $C_i(t,\mu)$,
depending on the (${\overline{\rm MS}}$) renormalization scale $\mu$ for 
the bilocal operator.
Apparently, the lowest-dimensional operator 
that participates in the RHS
is given by the dimension-3 operator appearing in the definition (\ref{fmu})
for the decay constant $F(\mu)$,
and, here, a complete set of local operators
of dimension $d = 4$ and 5 is also taken into account. 
The dimension counting tells us that the Wilson coefficients 
associated with the dimension-$d$ operators
behave as $C_i(t,\mu) \sim t^{d-3}$ modulo logarithm.
% as $t\rightarrow 0$.
Those coefficient functions are calculated to the NLO ($O(\alpha_s)$) accuracy,
and the corresponding NLO corrections turn out to induce 
the contributions associated with the logarithm $L$ of (\ref{L}).
Substituting the result for (\ref{ope0}) into (\ref{eq1}), 
the OPE form of the $B$-meson LCDA was derived as~\cite{Kawamura:2008vq},
\begin{eqnarray}
&& \!\!\!\!\!\!\!\!\!\!\!\!\!\!\!\!\!\!\!\!\!\!\!\!\!\!\!\!
\tilde{\phi}_+^{\rm OPE}(t,\mu)
=1- \frac{\alpha_s C_F}{4\pi}
\left(2L^2+2L+\frac{5 \pi^2}{12}\right)
-it\frac{4\bar{\Lambda}}{3} 
\left[1- \frac{\alpha_s C_F}{4\pi}
\left(2L^2+4L-\frac{9}{4}+\frac{5\pi^2}{12} \right)
\right]
\nonumber\\
&&
-t^2 \bar{\Lambda}^2 \!\! \left[
1\! - \! \frac{\alpha_sC_F}{4\pi}\!\!
\left(2L^2+\frac{16}{3}L-\frac{35}{9}
+\frac{5\pi^2}{12} \right)\!
\right]
\!\!- \! \frac{t^2\lambda_E^2(\mu)}{3}\!
\left[1\! - \! \frac{\alpha_sC_F}{4\pi}\!
\left( \! 2L^2+2L-\frac{2}{3}
+\frac{5\pi^2}{12} \! \right)
\right.
\nonumber\\
&&
\left.
+ 
\frac{\alpha_sC_G}{4\pi}
\left(\frac{3}{4}L-\frac{1}{2}\right)
\right]
-\frac{t^2\lambda_H^2(\mu)}{6} 
\left[1- \frac{\alpha_sC_F}{4\pi}
\left(2L^2+\frac{2}{3}
+\frac{5\pi^2}{12} \right)
-\frac{\alpha_s C_G}{8\pi}
\left(L-1\right)
\right]\ ,
\label{ope}
\end{eqnarray}
with $\alpha_s \equiv \alpha_s (\mu)$, as
usual.
This OPE form enables us to evaluate the $B$-meson LCDA
for interquark distances $t$ with $t \lesssim 1/\mu$
in a rigorous way
in terms of three nonperturbative parameters in the HQET, 
i.e., a familiar HQET parameter,
%one of which is 
as the mass difference between the $B$-meson and $b$-quark,
\be
\bar{\Lambda}=m_B -m_b\ ,
\label{lambdabar}
\ee
which is known to be associated with matrix elements of
dimension-4 operators~\cite{Neubert:1994mb}, 
and the novel HQET parameters $\lambda_E^2(\mu)$ and $\lambda_H^2(\mu)$,
which are defined by matrix elements of the 
quark-antiquark-gluon three-body operators of 
dimension~5 as~\cite{Grozin:1997pq,KKQT,GW}
\be
\langle 0|\bar{q}g \boldsymbol{E}\cdot \boldsymbol{\alpha}
\gamma_5h_v|B(v)\rangle=F(\mu)\lambda_E^2(\mu)\ ,~~~~~~~ 
\langle 0|\bar{q}g \boldsymbol{H}\cdot \boldsymbol{\sigma}
\gamma_5h_v|B(v)\rangle=iF(\mu)\lambda_H^2(\mu) \ ,
\label{lambdaeh}
\ee
associated with the chromoelectric and chromomagnetic fields, respectively, 
in the rest frame ($v=(1,{\bf 0})$).
We note that 
the ``universal'' double logarithmic term, $-(\alpha_s C_F/4\pi)2L^2$, in (\ref{ope})
yields as a UV-finite term from the contribution of the diagram of Fig.~\ref{fig:1}~(a)
%a  correction around the cusp,
in the one-loop matching calculation of the Wilson coefficients, 
%as , 
while the UV-divergent part from the same diagram 
%Fig.~\ref{fig:1}~(a)
induced the cusp anomalous
dimension in the evolution equation~(\ref{evol}) 
through the renormalization constant in (\ref{ren1}),
as discussed in Sec.~\ref{sec2}.
Indeed, 
the derivative of the double logarithmic terms in (\ref{ope})
with respect to $\mu$ reproduces the term associated with the cusp anomalous dimension
in (\ref{evol}).
Taking also the derivative of the other terms in (\ref{ope})
and 
combining the result with the scale dependence of the HQET parameters
of (\ref{lambdabar}) and (\ref{lambdaeh}),
i.e., $d \bar{\Lambda}/d\mu=0$ and that of
$\lambda^2_{E,H}(\mu)$ in terms of the
the one-loop mixing matrix obtained in \cite{GN97},
one can show that 
%the OPE form 
(\ref{ope}) 
satisfies the evolution equation 
(\ref{evol}),
as demonstrated in \cite{Kawamura:2008vq}.
Namely, 
the OPE form (\ref{ope}) 
for the LCDA
is completely 
consistent with the RG structure of (\ref{eq1})
embodied in (\ref{evol}).
This important property can be demonstrated
in an alternative way
using our NLL-level solution~(\ref{solution:4}) with (\ref{exponent})-(\ref{func:g3}):
expanding this solution 
in powers of $\alpha_s(\mu)$,
we obtain
\bea
&&\hspace{-1cm}
\tilde{\phi}_+(-i\tau,\mu)=
\left[1-\frac{\alpha_s C_F}{2\pi} \ln^2{\frac{\mu}{\mu_0}}
-\frac{\alpha_s C_F}{2\pi} \ln\frac{\mu}{\mu_0}
-\frac{\alpha_s C_F}{\pi} \ln(\tau\mu_0e^{\gamma_E}) \ln\frac{\mu}{\mu_0} 
\right] \tilde{\phi}_+(-i \tau, \mu_0)
\non\\
&&\hspace{1cm}
+\frac{\alpha_s C_F}{\pi} \ln\frac{\mu}{\mu_0}
\int_0^1 dz \tilde{\phi}_+( -i\tau z ,\mu_0)\left(\frac{z}{1-z}\right)_+ 
+\cdots\ ,
\label{LO}
\eea
where 
the anomalous dimensions (\ref{rgf1}), (\ref{rgf12}) are substituted,
and the ellipses stand for the terms of order $\alpha_s^2$ and
higher.
%Then, we 
Calculating the RHS of (\ref{LO}) 
%using 
with the substitutions,
$\tilde{\phi}_+ (-i\tau ,\mu_0) \rightarrow \tilde{\phi}_+^{\rm OPE} (-i\tau ,\mu_0)$,
$\tilde{\phi}_+ (-i\tau z,\mu_0) \rightarrow \tilde{\phi}_+^{\rm OPE} (-i\tau z,\mu_0)$,
we find that
the result to $O(\alpha_s)$
%taking into account the scale dependence of $\lambda^2_{E,H}(\mu)$ mentioned above,
reproduces 
exactly $\tilde{\phi}_+^{\rm OPE} (-i\tau ,\mu)$ given by (\ref{ope}).

From the above discussions, 
our formula~(\ref{solution:4}) with (\ref{exponent})-(\ref{func:g3})
and with the input DA as
$\tilde{\phi}_+ (-i\tau z,\mu_0) \rightarrow \tilde{\phi}_+^{\rm OPE} (-i\tau z,\mu_0)$ 
embodies the model-independent
form of $B$-meson LCDA~(\ref{eq1}), incorporating the RG improvement
%at the scale $\mu$, including 
associated with the NLL-level resummation, and
satisfying that its expansion to the fixed order $\alpha_s$ coincides with (\ref{ope}). 
%to coincide with (\ref{ope}) when expanded to the fixed order $\alpha_s$.
We emphasize that this form for the LCDA at the scale $\mu$ 
%($ > \mu_0$) 
is useful when $\tau \lesssim 1/\mu_0$:
recall that the OPE (\ref{ope}) with the local operators and the corresponding 
Wilson coefficients 
calculated in the fixed-order perturbation theory 
%associated with the local operators 
is useful 
when the interquark separation is less than the typical distance scale
of quantum fluctuation, i.e., $\tau \lesssim 1/\mu$~\cite{Kawamura:2008vq};
but, 
%as an important consequence of 
the quasilocal structure in (\ref{solution:4}), manifested as its integral
over $0 \le z \le 1$,
reveals that
this formula is applicable when the OPE form $\tilde{\phi}_+^{\rm OPE} (-i\tau z,\mu_0)$ 
substituted into the input DA 
is useful.
Namely, 
the range of applicability of the OPE (\ref{ope}) obtained in perturbation theory is extended
by performing the RG resummation as (\ref{solution:4}) to all orders in $\alpha_s$.
In the following, we choose the input scale as
$\mu_0\equiv 1$~GeV, corresponding to the typical hadronic scale,
and evaluate (\ref{solution:4}) with $\mu = \mu_{\rm hc} \sim \sqrt{m_b \Lambda_{\rm QCD}}$.

\begin{figure}
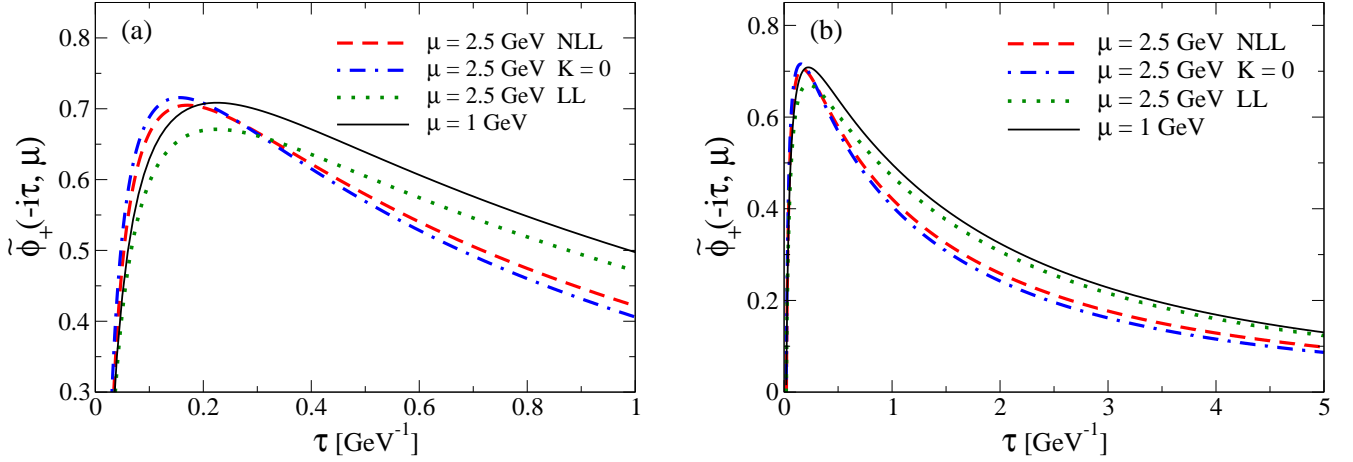

\hspace*{-0.5cm}
　\includegraphics[height=6.1cm]{fig2a.eps}~~~
　\includegraphics[height=6.1cm]{fig2b.eps}
\caption{The evolution of the $B$-meson LCDA based on the coordinate-space 
formula~(\ref{solution:4}):
%evolved from $\mu=1.0$ to $2.5$~GeV: 
the solid curve shows the input DA, 
%at $\mu=1$~GeV, 
given by (a) 
$\tilde{\phi}_+^{\rm OPE}(-i\tau,\mu_0)$ 
of (\ref{ope}) and (b) its extension 
using (\ref{input}).
The dashed, dot-dashed, and dotted curves show the results evolved to 
$\mu=2.5$~GeV,
using the evolution operator of (\ref{solution:4}) with the NLL accuracy,
using the evolution operator~(\ref{prefactor}) corresponding to $K=0$ in
(\ref{evol2}), and using the LL-level evolution, respectively.
}
\label{fig:2}
\end{figure}

We now specify the values of the HQET parameters 
participating in our input DA,
(\ref{ope}) with $\mu=\mu_0$.
For those parameters,
%them, 
we adopt the same input values as used in our previous work~\cite{Kawamura:2008vq}:
%we first note that 
$\bar{\Lambda}$ of (\ref{lambdabar}), which is defined by
the $b$-quark pole mass $m_b$,
%following \cite{Lee:2005gza}, we 
is eliminated 
%$\bar{\Lambda}$ 
in favor of a short-distance parameter, $\bar{\Lambda}_{\rm DA}$,
free from IR renormalon ambiguities~\cite{Bigi:1994} and written as
\be
\bar{\Lambda}
   = \bar{\Lambda}_{\rm DA}(\mu)
    \left[ 1 + \frac{7C_F\alpha_s (\mu)}{16\pi} \right] 
-\mu \frac{9C_F\alpha_s(\mu)}{8\pi}\ ,
\label{lambdada}
\ee
to one-loop accuracy. Here,
$\bar{\Lambda}_{\rm DA}(\mu)$ can be related to another short-distance mass parameter
whose value is extracted from analysis of the spectra in inclusive decays $B\to X_s\gamma$ 
and $B\to X_u l\,\nu$, leading to
$\bar{\Lambda}_{\rm DA}(\mu_0)\simeq 0.52$~GeV.
For the other two 
%novel 
HQET parameters,
% associated with the dimension-5 operators, 
we use the central values of
\begin{equation}
\lambda_E^2 (\mu_0) = 0.11 \pm 0.06~{\rm GeV}^2\ , \;\;\;\; \;\;\;\;\;\;\;\;
\lambda_H^2 (\mu_0) = 0.18 \pm 0.07~{\rm GeV}^2\ ,
\label{lambdaEH}
\end{equation}
%at $\mu =1$~GeV,
which were 
obtained by QCD sum rules~\cite{Grozin:1997pq}; at present, no other estimate
exists for $\lambda_{E}^2$ or $\lambda_{H}^2$.
We now calculate
(\ref{ope}) with $\mu=\mu_0$
and the imaginary light-cone separation
%performing the Wick rotation
as $t\rightarrow -i\tau$,
and obtain
model-independent description of 
the $B$-meson LCDA
$\tilde{\phi}_+^{\rm OPE}(-i\tau,\mu_0)$ for 
%small 
$\tau \lesssim 1/\mu_0=1$~GeV$^{-1}$,
%The initial DA (\ref{ope}) at $\mu=1$GeV 
which is displayed by the solid line in Fig.~\ref{fig:2}~(a). 
This result can be substituted
directly into the RHS of (\ref{solution:4}) as the input LCDA
for the case with $\tau \lesssim 1$~GeV$^{-1}$, 
because $z\tau\le \tau \lesssim 1$~GeV$^{-1}=1/\mu_0$
in the integrand, reflecting the quasilocal nature as noted above.
%%%%%%%%%%%%%%%%%%%%%%%%%%%%%%%%%%%%%%%%%%%%%%%%%%%%%%%%%%%%%%%%%%%%%%%%%%%%%%%
%%%%%%%%%%%%%%%%%%%%%%%%%%%%%%%%%%%%%%%%%%%%%%%%%%%%%%%%%%%%%%%%%%%%%%%%%%%%%%%

We now discuss the results of our evolution (\ref{solution:4}) 
to higher scale $\mu \sim \sqrt{m_b \Lambda_{\rm QCD}}$,
shown in Fig.~\ref{fig:2}~(a):
the dashed line 
denotes the full result of the LCDA $\tilde{\phi}_+(-i\tau,\mu)$ at $\mu=2.5$~GeV, 
obtained by our
NLL evolution (\ref{solution:4}) using (\ref{exponent})-(\ref{func:g3}).
When we omit the effect of 
the DGLAP-type kernel $K(z, \alpha_s)$ in the evolution equation (\ref{evol2}),
the resultant evolution is induced only by the factor (\ref{prefactor})
(see (\ref{U2})),
yielding the dot-dashed curve.
Omitting the other NLL terms in (\ref{solution:4}) further, as 
${\cal V}(\mu,\mu_0)\rightarrow (4\pi/\alpha_s(\mu))h^{(0)}(\chi)$
and $\xi \rightarrow 0$,
we obtain the dotted curve that corresponds to the result of the LL-level evolution. 
We see the considerable Sudakov suppression not only at the LL level
but also at the NLL level; in particular, at the NLL level, the suppression
arises in the moderate $\tau$ regions, 
%$ \tau \gtrsim 1/\mu$,
while the DA is enhanced for small $\tau$,
reflecting the $\tau$ dependence of the factor (\ref{prefactor}).
On the other hand, the DGLAP-type kernel contributes to shifting the distribution 
from small to moderate $\tau$, as a result of the integral over $z$ in (\ref{solution:4});
such effect is  
characteristic of the evolution that is induced by the 
kernel associated with the plus-distribution of the type (\ref{plusd}),
and is similar to
the corresponding effects arising in the usual DGLAP equation
for the parton distribution functions of the nucleon.
From the discussion above (\ref{lambdada}), our full result, the dashed curve,
is useful for providing model-independent behavior of the $B$-meson LCDA
in small and moderate $\tau$ regions presented in Fig.~\ref{fig:2}~(a),
where the solid curve using the OPE form $\tilde{\phi}_+^{\rm OPE}(-i\tau,\mu_0)$ 
is suitable for the input DA.

The OPE form (\ref{ope}), used for the input DA, breaks down
in the large $\tau$ region, 
where the contributions associated with the operators
of any higher dimension become important because
the contributions from the dimension-$d$ operators grow as 
$\sim \tau^{d-3}$ with increasing $\tau$.
According to our previous work~\cite{Kawamura:2008vq},
we rely on a model function to describe the DA in the large $\tau$ region
dominated by the nonperturbative effects, and,
specifically, we use the following form of the input DA 
for the entire range of $\tau$,
\bea
\tilde{\phi}_+(-i\tau,\mu_0)=\theta(\tau_c-\tau) 
\tilde{\phi}_+^{\rm OPE}(-i\tau,\mu_0)
+\theta(\tau-\tau_c) \frac{N}{(\tau\omega_0+1)^2}\ ,
\label{input}
\eea
with $\tau_c \sim 1/\mu_0$, such that
we connect the small and moderate $\tau$ behavior given by the rigorous OPE form,
the first term,
smoothly to a model for the large $\tau$ behavior, the second term.
This second term corresponds to the exponential form 
$N(\omega/\omega_0^2)e^{-\omega/\omega_0}$ 
in the momentum representation, 
and such form was
suggested in an estimate of the $B$-meson LCDA 
using QCD sum rules~\cite{Grozin:1997pq}
and was also adopted in~\cite{Lee:2005gza} as a nonperturbative component
to model the $B$-meson LCDA 
using the information from 
the OPE with the local operators of dimension $d \le 4$
and the NLO corrections to the corresponding Wilson coefficients taken into account.
(For the correspondence and difference between our OPE~(\ref{ope})
and the OPE derived in~\cite{Lee:2005gza}, see the discussion in \cite{Kawamura:2008vq}.)
Here, the two parameters $N$ and $\omega_0$ 
are determined by the continuity of (\ref{input}) and its derivative,
$\tilde{\phi}_+(-i\tau,\mu_0)$ and 
$\partial \tilde{\phi}_+(-i\tau,\mu_0)/\partial \tau$,
at $\tau=\tau_c$. 
The resulting values $N\simeq0.86$ and $\omega_0\simeq0.31$~GeV
are found to be stable under the variation of $\tau_c$ 
for $0.6~{\rm GeV}^{-1} \lesssim \tau_c \lesssim 1~{\rm GeV}^{-1}$,
and so is the behavior 
of the corresponding DA (\ref{input})~\cite{Kawamura:2008vq}.
In the following, we take $\tau_c=1$~GeV$^{-1}$, and now the solid curve 
in Fig.~\ref{fig:2}~(a) is continued to the $\tau \ge 1$~GeV$^{-1}$ region 
with (\ref{input}),
as presented by the solid curve in Fig.~\ref{fig:2}~(b).
Using this result of (\ref{input}) as the input DA in (\ref{solution:4}),
we obtain the other curves in Fig.~\ref{fig:2}~(b), which are evolved 
in the same way as
the corresponding curves in Fig.~\ref{fig:2}~(a); in particular,
the behaviors of those new curves in the region $\tau \le \tau_c=1$~GeV$^{-1}$ 
completely coincide with those of
the corresponding curves in Fig.~\ref{fig:2}~(a).
Namely, the model-independent nature 
%of the input distribution (\ref{input}) 
for $\tau \le \tau_c$,
%based on 
originating from the OPE,
is preserved under the evolution.
This remarkable feature of our results is a direct consequence of the 
quasilocal structure of the evolution~(\ref{solution:4}) 
in the coordinate-space representation: 
the results
in the region $\tau \le \tau_c$ are not contaminated under the evolution by
the assumed model behavior for larger distances, 
the contribution of the second term of (\ref{input}).
On the other hand, the smooth continuation of this second term of (\ref{input}) to the
first term at $\tau =\tau_c$ and the evolutions of the result
lead to the similar interrelations between the curves at
large $\tau >\tau_c$ in Fig.~\ref{fig:2}~(b) as those at moderate $\tau$, displayed also
in Fig.~\ref{fig:2}~(a); e.g., we observe the considerable Sudakov suppression also
at large $\tau$ region.

Integrating the dashed curve in Fig.~\ref{fig:2}~(b) over the entire range of $\tau$, 
we obtain the first inverse moment of the $B$-meson LCDA (see (\ref{sigman})-(\ref{lsr})),
\begin{equation}
\lambda_B^{-1}(\mu) 
= \int_0^{\tau_c} d\tau 
\tilde{\phi}_{+}(-i\tau, \mu)+ \int_{\tau_c}^\infty d\tau 
\tilde{\phi}_{+}(-i\tau, \mu)\ ,
\label{lambdaB}
\end{equation}
at the scale $\mu=2.5$~GeV,
as the sum of the model-independent contribution, the first term, originating from the OPE
and the second term depending on the assumed model behavior at large distances.
Table~\ref{tab:1} shows the results for $\mu=2.5$~GeV and some other values of $\mu$,
with the first and second numbers in the parentheses
denoting the contributions from the first and second terms in~(\ref{lambdaB}).
One can check that the exactly same values of $\lambda_B^{-1}(\mu)$ are obtained 
using the formula~(\ref{lambda_B}),
as the integrals of 
the input DA (\ref{input})
with the corresponding weight function
at the NLL accuracy.
\begin{table}
\begin{center}
\begin{tabular}{|c|c|c|c|}
\hline
%$\mu$ [GeV] 
&\multicolumn{3}{|c|}{$\lambda_B^{-1}(\mu)$~{\small [GeV$^{-1}$]}
}
\\
\hline
$\mu$~{\small [GeV]}& Eq.~(\ref{lambdaB}) with Eqs.~(\ref{solution:4}), (\ref{input}) 
& Lee-Neubert
& Braun {\it et al.}
\\
\hline
1.0  & $2.7\, \  ( 0.6  + 2.1  )$ & 2.1& 2.2\\
1.5  & $2.4\, \  ( 0.6  + 1.8  )$ & 1.9 &2.0 \\
2.0  & $2.2\, \  ( 0.5  + 1.7 )$ & 1.7 & 1.9\\
2.5  & $2.1\, \  ( 0.5  + 1.6  )$ & 1.6 & 1.8 \\
\hline
\end{tabular}
\end{center}
\caption{
The results of the inverse moment $\lambda_B^{-1}(\mu)$ 
%obtained by the formula~of (\ref{lambdaB}) 
using (\ref{solution:4}) to the NLL accuracy for the input DA~(\ref{input}),
% $\mu=1$~GeV, 
with the first and second 
numbers in the parentheses denoting the contributions from the first and the second terms
in (\ref{lambdaB}).
The results obtained by Lee and Neubert~\cite{Lee:2005gza} 
and the estimates based on the calculation 
%at $\mu=1$~GeV 
by Braun {\it et al.}~\cite{Braun:2003wx} are also shown for comparison. 
}
\label{tab:1}
\end{table}
In Table~\ref{tab:1}, we note that the result for $\mu=\mu_0=1.0$~GeV coincides with that reported in
our previous work~\cite{Kawamura:2008vq}. The evolution decreases 
$\lambda_B^{-1}(\mu)$ 
with increasing $\mu$,
in particular, through the decrease of the model-dependent contribution, 
the second term of (\ref{lambdaB}).
On the other hand, our results of $\lambda_B^{-1}(\mu)$ 
for $\mu \sim \sqrt{m_b \Lambda_{\rm QCD}}$
are larger than the results of \cite{Lee:2005gza} 
as well as of \cite{Braun:2003wx},
where, for the former case in Table~\ref{tab:1}, we quote the results calculated 
in \cite{Lee:2005gza}, and, for the latter case, we present estimates 
with the fixed-order formula~(\ref{lambda_B:LO})
substituting $\lambda_B^{-1}(\mu_0)$ and $\sigma_1(\mu_0)$ (see (\ref{lsr}))
that were obtained 
in~\cite{Braun:2003wx}.
We recognize that the evolution from $\mu= \mu_0$ 
to $\mu_{\rm hc} \sim \sqrt{m_b \Lambda_{\rm QCD}}$ could give rise
to the decrease of $\lambda_B^{-1}(\mu)$ by 20-30\%,
with the larger value of $\lambda_B^{-1}(\mu_0)$ leading to the
larger $\lambda_B^{-1}(\mu_{\rm hc})$; 
%for $\mu \sim \mu_{\rm hc}$; 
as emphasized 
in \cite{Kawamura:2008vq},
our larger $\lambda_B^{-1}(\mu_0)$ than 
the corresponding values of
other works~\cite{Lee:2005gza,Braun:2003wx}  
originates from the novel contribution of $\lambda_E^2$ and $\lambda_H^2$ in 
the OPE form (\ref{ope}),
which are associated with the dimension-5 operators 
representing the quark-antiquark-gluon three-body correlation (see (\ref{lambdaeh})).

%%%%%%%%%%%%%%%%%%%%%%%%%%%%%%%%%%%%%%%%%%%%%%%%%%%%%%%%%%%%%%%%%%%%%%%%%%%%%%%
\begin{figure}
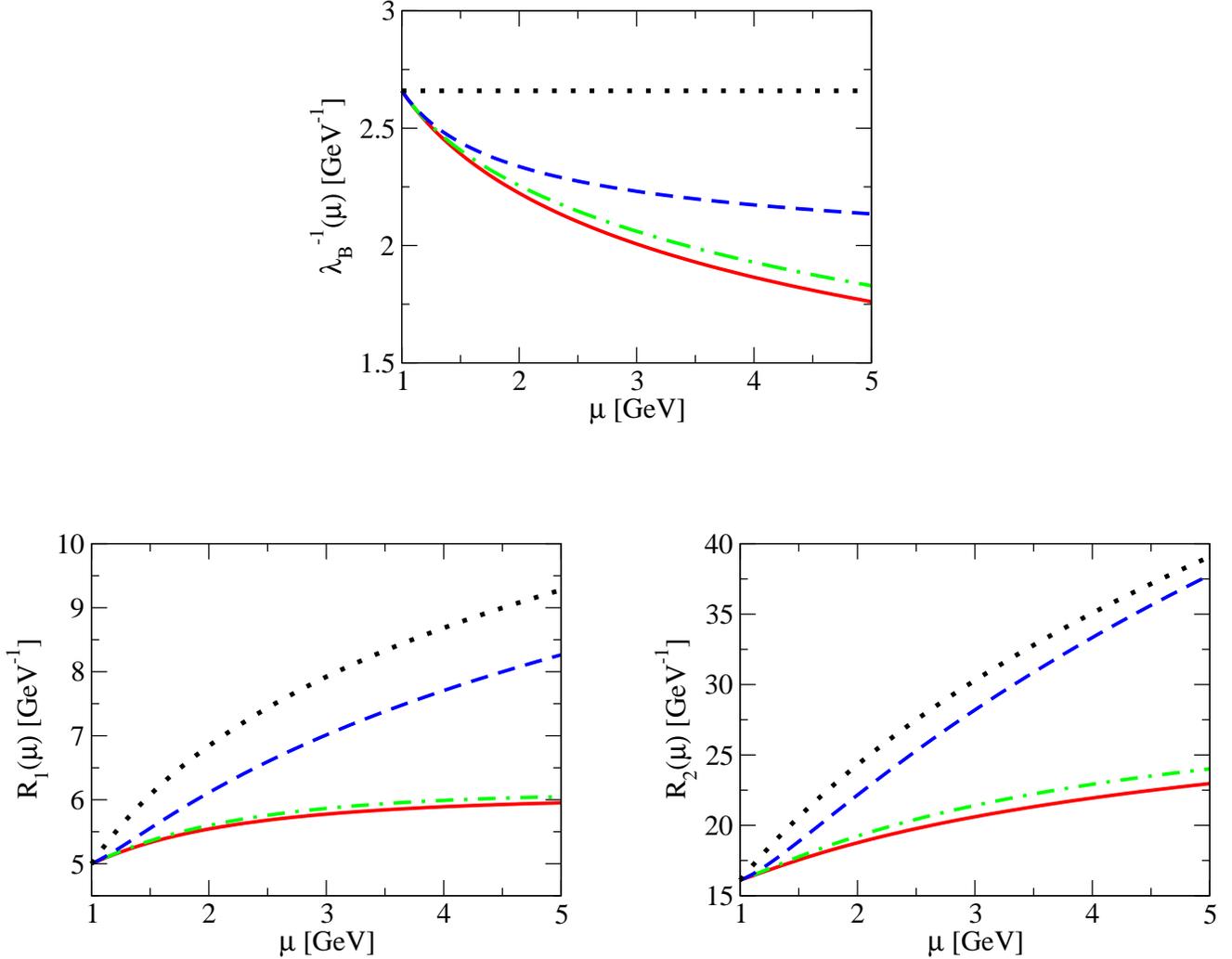

\hspace{-1cm}
　\includegraphics[height=6.1cm]{fig3lambda.eps}\\
\vspace{1.5cm}
\hspace{-0.5cm}
　\includegraphics[height=6.1cm]{fig3R1.eps}~~~~~~~~~~
  \includegraphics[height=6.1cm]{fig3R2.eps}
\caption{The evolution of the first few logarithmic moments defined in the coordinate 
space, (\ref{lnmom}), using (\ref{input}) as the input $B$-meson LCDA 
at the scale $\mu_0=1$~GeV:
%evolved from $\mu=1.0$ to $2.5$~GeV: 
the solid curves show the results of (\ref{lambda_B})-(\ref{R2}) at 
the NLL accuracy,
the dot-dashed curves show  
the fixed-order results (\ref{lambda_B:LO})-(\ref{R2:LO}),
and the dotted curves show the results when $\alpha_s \rightarrow 0$.
The dashed curves show the fixed-order results (\ref{lambda_B:LO})-(\ref{R2:LO})
with 
the dominant double-logarithmic corrections omitted.
}
\label{fig:3}
\end{figure}
\begin{figure}
\bc
　\includegraphics[height=6.1cm]{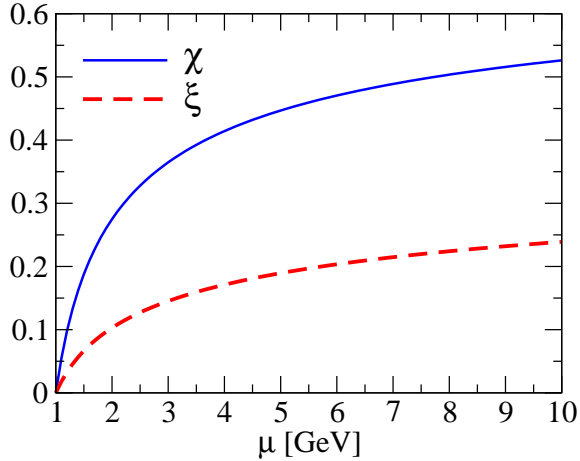}
\ec
\caption{$\chi$ of (\ref{lambda}) and $\xi$ of (\ref{func:g3})
as functions of $\mu$ with $\mu_0=1$~GeV, 
shown by the solid and dashed curves, respectively.}
\label{fig:4}
\end{figure}

%%%%%%%%%%%%%%%%%%%%%%%%%%%%%%%%%%%%%%%%%%%%%%%%%%%%%%%%%%%%%%%%%%%%%%%%%%%%%%%
Our results of $\lambda_B^{-1}(\mu)$ presented in Table~\ref{tab:1}
as well as those calculated at even higher $\mu$
are plotted 
by the solid curve in the first panel in Fig.~\ref{fig:3},
and, similarly, the solid curves in the other two panels
show the behaviors of the logarithmic moments 
defined in the coordinate space,
$R_1(\mu)$ and $R_2(\mu)$ of (\ref{R1}) and (\ref{R2}),
with the NLL accuracy (\ref{exponent})-(\ref{func:g3}) using the input DA (\ref{input}).
%by the solid curves in the other panels.
Here, the dot-dashed curves present 
the fixed-order calculations based on (\ref{lambda_B:LO})-(\ref{R2:LO})
substituting $\lambda_B^{-1}(\mu_0)$, $R_1(\mu_0)$, $R_2(\mu_0)$ and $R_3(\mu_0)$ 
calculated with (\ref{input}),
and these results are modified into the dashed curves 
when we omit the double logarithmic 
correction behaving as $\propto \alpha_s\ln^2 (\mu/\mu_0)$,
compared to the corresponding tree ($O(\alpha_s^0)$) contribution,
in each coefficient of 
$\lambda_B^{-1}(\mu_0)$, $R_1(\mu_0)$, $R_2(\mu_0)$ and $R_3(\mu_0)$
in the RHS of (\ref{lambda_B:LO})-(\ref{R2:LO}).
Furthermore, those results reduce to the dotted lines 
%Also, the dotted lines show the results of (\ref{lambda_B:LO})-(\ref{R2:LO})
when $\alpha_s \rightarrow 0$.
We find good accuracy of the fixed-order formulae 
(\ref{lambda_B:LO})-(\ref{R2:LO}),
and thus the rapid convergence of the resummed perturbation theory in 
(\ref{lambda_B})-(\ref{R2}) with (\ref{exponent})-(\ref{func:g3}) 
when organized by $\chi$ of (\ref{lambda}), whose behavior
as a function of $\mu$ is shown by the solid curve in Fig.~\ref{fig:4}.
On the other hand, in (\ref{lambda_B:LO})-(\ref{R2:LO}), the double logarithmic effects
play important roles to determine the scale dependence of  
$\lambda_B^{-1}(\mu)$, $R_1(\mu)$, $R_2(\mu)$,
while the other $O(\alpha_s)$ 
contributions tend to cancel to a large extent.
As the result, the perturbative evolution from $\mu=\mu_0$ to 
$\mu_{\rm hc}$
can modify the values of those logarithmic
moments considerably, by 20-30\%.
In Fig.~\ref{fig:4}, we also show the behavior of $\xi$ of (\ref{func:g3})
by the dashed curve; this demonstrates that the condition~(\ref{xiband2})
is indeed satisfied, so that 
our formulae (\ref{lambda_B})-(\ref{R2}), as well as
(\ref{solution:4}) giving their basis, 
describe the well-defined evolutions for all relevant scales.

In the present paper, we have discussed in detail 
the effects of the evolution on the $B$-meson LCDA
and on its integrals $\lambda_B^{-1}(\mu)$, $R_1(\mu)$, $R_2(\mu)$ relevant to 
exclusive $B$ decays,
emphasizing model-independent aspects
revealed by our coordinate-space approach,
but did not intend to determine the precise values of 
those integrals 
at $\mu = \mu_{\rm hc}$.
Such determination requires us to calculate 
the $B$-meson LCDA~(\ref{eq1}) at the initial scale $\mu_0$,
reducing the corresponding theoretical uncertainty as much as possible:
as found in our previous work~\cite{Kawamura:2008vq},
the corresponding initial LCDA, calculated in the form of (\ref{input}),
is significantly influenced by
the novel
HQET parameters $\lambda_E$ and $\lambda_H$
arising in 
the OPE form (\ref{ope}),
which are 
associated with the dimension-5 quark-antiquark-gluon operators.
Therefore, the rather large uncertainty in their existing estimate~(\ref{lambdaEH})
based on QCD sum rules
%This finding 
calls for 
more precise 
%nonperturbative
estimates of 
$\lambda_E$ and $\lambda_H$.
Recently, higher-order corrections to the QCD sum rules for $\lambda_E$ and $\lambda_H$
have been calculated, and these new contributions are found to improve 
the estimate of $\lambda_E$ and $\lambda_H$~\cite{nisi}.
We also note 
that in the RHS of (\ref{lambdaB}), evaluated in Table~\ref{tab:1},
the second term is much larger than the first term.
This suggests that $\lambda_B$, as well as $R_{1,2}$, is rather sensitive to 
the functional form
that models 
the LCDA in the long-distance region; for example,
a functional form motivated by the so-called Wandzura-Wilczek 
approximation~\cite{KKQT,GW}
provides an interesting possible 
alternative to the form appearing in the second term in (\ref{input})
(see, e.g., \cite{Huang:2005kk} for other studies
on the behaviors of the LCDA).
Systematic investigations of these points,
combined with the evolution effects obtained in this paper,
could determine the values of $\lambda_B^{-1}(\mu)$, $R_1(\mu)$, $R_2(\mu)$ 
at $\mu= \mu_{\rm hc}$ as precisely as possible, and
%with the error estimation, and 
those results will be presented elsewhere.

\section{Conclusions}
\label{sec6}

In this paper, we have studied the RG evolution of the $B$-meson LCDA,
working in the coordinate-space representation 
of the LCDA.
The corresponding evolution equation and its solution
%, derived in this paper,
demonstrated that our coordinate-space approach has remarkable advantages over 
the conventional approach in the momentum space.
Indeed, only in the coordinate space, the relevant kernel in the evolution equation,
associated with the cusp anomalous dimension as well as the DGLAP-type 
anomalous dimension, is quasilocal, and this quasilocality is
inherited by the corresponding analytic solution,
leading to the simplest expression possible for calculating the evolution of the 
$B$-meson LCDA.
Our explicit formula of the solution has the accuracy beyond the 
one-loop level in the RG-improved perturbation theory,
%and gives the NLL-resummed evolution 
taking into account 
the effect of the two-loop cusp anomalous dimension 
according to consistent order counting,
such that
%where 
the Sudakov-type double logarithmic effects as well as the DGLAP-type 
single-logarithmic corrections are resummed at the NLL accuracy. 
This result, in turn, allowed us to 
derive the master formula, by which
the relevant integrals of the LCDA
at the scale $\mu_{\rm hc}\sim\sqrt{m_b \Lambda_{\rm QCD}}$, arising in 
the factorization formula for the exclusive $B$-meson decays,
can be reexpressed in a model-independent way
by the compact integrals of the LCDA at a typical hadronic scale $\mu_0 \sim 1$~GeV.

We applied our evolution formula to the LCDA with the initial scale $\mu_0$,
which is determined by the OPE having the perturbative (NLO) accuracy 
consistent with the NLL-level resummation, and the highest nonperturbative accuracy,
at present,
taking into account the local operators of dimension $d \le 5$.
The quasilocal structure of our evolution guarantees that
the $B$-meson
LCDA at a certain quark-antiquark distance
is not contaminated under the change of the renormalization scale 
by 
%the uncertainty in 
the configurations of the quark and antiquark 
for the larger distances,
% based on the OPE
so that the LCDA at high scales, obtained through our evolution
from the OPE-based, initial LCDA that is accurate for interquark distances
less than $\sim 1/\mu_0\sim 1$~GeV$^{-1}$,
exhibits 
%determines 
the model-independent behaviors for distances $\lesssim 1$~GeV$^{-1}$.
Our explicit numerical calculation indicated
the considerable effects of the evolution, from the scale $\mu_0$
to $\mu_{\rm hc}$,
for the LCDA 
and its relevant integrals.
In particular, we found that the dominant roles are
played by the double logarithmic corrections,
although particular attention was not paid to them in previous works.
On the other hand, we observed the rapid convergence
of the corresponding resummed perturbation series organized
by the proper logarithmic expansion.

Using the information available 
%at present 
for the nonperturbative effects associated with the OPE-based, initial LCDA,
our evolution gave an estimate for the relevant integrals of the LCDA at
the scale $\mu_{\rm hc}$,
e.g.,
$\lambda_B^{-1}(\mu)\simeq 2.1$~GeV$^{-1}$ at $\mu= 2.5$~GeV.
This is 
%somewhat 
larger than 
%other known 
the estimates by other works,
inheriting the larger value 
$\lambda_B^{-1}(\mu=1~{\rm GeV})\simeq 2.7$~GeV$^{-1}$
in our case, which is 
induced by matrix elements of the dimension-5 quark-antiquark-gluon operators 
in the OPE for the initial LCDA.
Combined with an update of the information on
%the relevant 
the nonperturbative effects in the initial LCDA,
the results derived in this paper are immediately applicable
for calculating the refined values of those integrals relevant to exclusive $B$ decays.

\appendix

\section{The evolution in the momentum representation}
\label{appa}

The evolution of the $B$-meson LCDA $\tilde{\phi}_+ (t, \mu)$
in the representation with the light-cone separation $t$
is provided by our solution (\ref{solution:4})
with the analytic continuation 
$\tau \rightarrow i(t-i0)$
performed.
We calculate 
the Fourier transformation of this result,
in order to derive the evolution for the LCDA $\phi_+(\omega, \mu)$
in the momentum representation
(see (\ref{eq1})):
\bea  
&&\hspace{-1cm}
\phi_+(\omega,\mu)
=\frac{1}{2\pi} \int_{-\infty}^{\infty} dt e^{i\omega t}\tilde{\phi}_+(t, \mu)
=\frac{e^{{\cal V}(\mu,\mu_0)+(1-2\gamma_E)\xi}}{\Gamma(\xi)\mu_0^{\xi}}
\int_0^\infty d\omega^\prime 
I(\omega,\omega^\prime)
\phi_+(\omega^\prime,\mu_0)\ ,
\label{FT:1}
\eea
with the integration kernel,
\be
I(\omega,\omega^\prime)
=\int_0^1dz \left(\frac{z}{1-z}\right)^{1-\xi}
\int_{-\infty}^{\infty} 
\frac{dt}{2\pi} e^{i(\omega-\omega^\prime z ) t} 
\left[i(t-i0)\right]^{-\xi}\ ,
\label{kernel:01}
\ee
where the integration over $t$ can be performed straightforwardly, yielding
\bea
&&
%\hspace{1cm}
I(\omega,\omega^\prime)
=\frac{1}{\Gamma(\xi)}
\int_0^{1
%{\rm min}\{1,\omega/\omega^\prime\}
}dz 
\left(\frac{z}{1-z}\right)^{1-\xi}
\frac{\theta(\omega-\omega^\prime z)}{(\omega-\omega^\prime z)^{1-\xi}}
\non\\
&&\hspace{1.5cm}
=\frac{1}{\Gamma(\xi)}
\int_0^{\frac{\omega_<}{\omega^\prime}
%{\rm min}\{1,\omega/\omega^\prime\}
}dz 
z^{1-\xi}(1-z)^{\xi-1}(\omega-\omega^\prime z )^{\xi-1}\ .
%\frac{1}{( \omega-\omega^\prime z )^{1-\xi}} \ .
\label{kernel:1}
\eea
Here, we have introduced the notation,
$\omega_< \equiv {\rm min}(\omega,\omega^\prime)$.
Changing the integration variable to $u=(\omega^\prime/\omega_<)z$
and using $\omega \omega^\prime = \omega_{>} \omega_{<}$
with $\omega_{>} \equiv{\rm max} (\omega,\omega^\prime)$,
we can rewrite 
%the kernel 
(\ref{kernel:1}) as
\be
%&&
I(\omega,\omega^\prime)=
\frac{\omega_>^{\xi-1}\omega_<}{\Gamma(\xi)\ \omega^\prime}
%\cdot\frac{1}{\Gamma(\xi)}
\int_0^1du u^{1-\xi}
\left(1- u\right)^{\xi-1}
\left(1-\frac{\omega_<}{\omega_>}u\right)^{\xi-1}\ ,
\label{kernel:2}
\ee
and we note that this can be expressed by the 
%where the 
hypergeometric function,
\bea
&&
{}_2F_1(\alpha,\beta;\gamma;z)\equiv
\frac{\Gamma(\gamma)}{\Gamma(\alpha)\Gamma(\beta)}
\sum_{n=0}^\infty
\frac{\Gamma(\alpha+n)\Gamma(\beta+n)}{\Gamma(\gamma+n)}
\frac{z^n}{n!}
\\&&\hspace{2.65cm}
=\frac{\Gamma(\gamma)}{\Gamma(\beta)\Gamma(\gamma-\beta)}
\int_0^1 du ~u^{\beta-1}(1-u)^{\gamma-\beta-1}(1-zu)^{-\alpha}\ ,
\eea
where the first line shows the usual definition by the series expansion,
and the second line gives the integral representation to be compared with
(\ref{kernel:2}).
Substituting the result into (\ref{FT:1}), we 
%finally 
obtain 
\be
\phi_+(\omega,\mu)=e^{{\cal V}(\mu,\mu_0)+(1-2\gamma_E){\cal \xi}}
\frac{\Gamma(2-\xi)}{\Gamma(\xi)}\!
\int_0^\infty \frac{d\omega^\prime}{\omega^\prime}
\phi_+(\omega^\prime,\mu_0)\left(\frac{\omega_>}{\mu_0}\right)^{\xi}
\frac{\omega_<}{\omega_>}~
{}_2 F_1\left(1-\xi,2-\xi;2;\frac{\omega_<}{\omega_>}\right)\! ,
%\non\\
\label{solution:mom}
\ee
which gives a well-defined formula when the condition (\ref{xiband}) is satisfied.
The evolution of the $B$-meson LCDA in this form
was first derived in \cite{Lange:2003ff,Lee:2005gza} 
by solving the evolution equation given 
in the momentum space, see (\ref{ft1}), (\ref{ft2}); note that
%\footnote{
${\cal V}(\mu,\mu_0)+\xi$ and 
$\xi$ in the present paper correspond, respectively, to  
$V(\mu,\mu_0)$ and $g$ in \cite{Lange:2003ff,Lee:2005gza}.
We mention that 
it would not be straightforward 
to derive the relations (\ref{integral3})-(\ref{sigma2})
based on (\ref{solution:mom}),
because of the structure involving the complicated integration of the 
hypergeometric function (see \cite{Bell:2008er}).

%%%%%%%%%%%%%%%%%%%%%%%%%%%%%%%%%%%%%%%%%%%%%%%%%%%%%%%%%%%%%%%%%%%%%%%%%%%%%%%

\section*{Acknowledgments}
\bigskip
We thank V.~M. Braun for valuable discussions. 
This work was supported by the Grant-in-Aid for Scientific Research 
No.~B-19340063. The work of H.K. is supported in part by the UK Science \& 
Technology Facilities Council under grant number PP/E007414/1.

\end{document}